\documentclass[conference]{IEEEtran}
\IEEEoverridecommandlockouts

\usepackage{moreverb,url}
\usepackage{amssymb,amsmath,amsfonts}
\usepackage{subfig}
\usepackage{graphicx}
\usepackage{rotating}
\usepackage{geometry}
\usepackage{setspace}
\usepackage{lipsum}
\newcommand{\verbatimfont}[1]{\renewcommand{\verbatim@font}{\ttfamily#1}}
\usepackage{epstopdf}
\usepackage{alltt}
\usepackage{color}
\usepackage{xcolor}
\usepackage{multirow}
\usepackage{listings}

\usepackage{fancyvrb}
\fvset{fontsize=\scriptsize}
\RecustomVerbatimEnvironment{verbatim}{Verbatim}{}

\usepackage{floatrow}
\usepackage{algpseudocode,algorithm,algorithmicx}
\usepackage{algcompatible}

\usepackage{algorithm}
\usepackage{algpseudocode}
\usepackage{cleveref}
\usepackage{booktabs}

\usepackage[frozencache,cachedir=.]{minted}

\newcommand{\figratio}{0.31}

\def\scriptO{{{\it O}\kern -.42em {\it `}\kern + .20em}}
\def\RR{{{\rm l}\kern - .15em {\rm R} }}
\def\PP{{{\rm l}\kern - .15em {\rm P} }}
\def\L2{{{\sf L}^2}}
\def\H1{{{\sf H}^1}}
\def\PN2{{\PP_{N}-\PP_{N-2}}}

\def\complex{{{\rm C} \kern - .53em {\rm l} \kern + .38em}}

\def\a1{{ | \lambda_{\min} |}}

\def\l1{{   \lambda_{\min}  }}

\def\bu0{{\underline {\bf 0}}}

\def\bu{{\bf u}}

\def\u0{{\underline 0}}
\def\1u{{\underline 1}}

\newcommand{\nomp}{{\em nomp}}
\newcommand{\nompcc}{{\em nompcc}}
\newcommand{\libnomp}{{\em libnomp}}

\begin{document}
\title{nomp: A Framework for Building Domain Specific Compilers}

\author{
\IEEEauthorblockN{
Thilina Ratnayaka\IEEEauthorrefmark{1}\IEEEauthorrefmark{2},
Kaushik Kulkarni\IEEEauthorrefmark{1}\IEEEauthorrefmark{3},
Nipuna Fernando\IEEEauthorrefmark{4},
Pubudu Hewavitharana\IEEEauthorrefmark{4},\\
Hirumal Priyashan\IEEEauthorrefmark{4},
Poorna Gunathilaka\IEEEauthorrefmark{4},
Nagitha Abeywickrema\IEEEauthorrefmark{4},
Ravindu Hirimuthugoda\IEEEauthorrefmark{4},\\
Tarun Prabhu\IEEEauthorrefmark{5},
Kirshanthan Sundararajah\IEEEauthorrefmark{6},
and Sanath Jayasena\IEEEauthorrefmark{4}
}
\IEEEauthorblockA{\IEEEauthorrefmark{1}Department of Computer Science, University of Illinois, Urbana, IL 61801}
\IEEEauthorblockA{\IEEEauthorrefmark{2}Argonne National Laboratory, Lemont, IL 60439\\
Email: tratnayaka@anl.gov}
\IEEEauthorblockA{\IEEEauthorrefmark{3}Qualcomm Research, San Diego, CA}
\IEEEauthorblockA{\IEEEauthorrefmark{4}Department of Computer Science, University of Moratuwa, Katubedda 10400}
\IEEEauthorblockA{\IEEEauthorrefmark{5}Los Alamos National Laboratory, Los Alamos, NM 87545}
\IEEEauthorblockA{\IEEEauthorrefmark{6}Department of Computer Science, Virginia Tech, Blacksburg, VA 24061}
}

\maketitle

\begin{abstract}
The low-level GPU programming models (CUDA, HIP, OpenCL, etc.)
provide detailed control of the data flow and execution plan of a
program in order to extract close-to-metal performance.
However, these have a steep learning curve due to the intricacies of
their syntax and semantics. This reduces programmer productivity.
On the other hand, high-level models (OpenMP, OpenACC, etc.)
that serve as abstractions over the low-level models are aimed at improving
programmer productivity but achieving performance on-par with the
low-level models is a challenge.
There are inherent trade-offs between productivity, portability and
performance in both approaches and there is no one-size-fits-all solution
which achieves all three simultaneously.
However, we believe there is room to improve programmer productivity without
sacrificing performance and portability by reusing optimization patterns
specific to a given domain.
To this end, we propose \nomp: a framework for building domain specific
compilers. \nomp~consists of a pragma based programming model and a runtime
capable of code transformation and generation based on user provided metadata.
\end{abstract}


\section{Introduction}\label{sec:intro}
Graphic processing units (GPUs) have emerged as the primary workhorse for
traditional high-performance computing (HPC) and modern artificial intelligence
(AI) applications in state-of-the-art computing systems.
This is evident by the fact that 9 of the top 10 systems in the latest
(June, 2024) edition of the TOP500 list are GPU-based
supercomputers~\cite{top500_2024_06}.
The number of GPU-based systems in the TOP500 list are growing fast and
already account for the majority of performance.
The Exascale Computing Project (a joint project funded by Department of Energy
Office of Science and National Nuclear Security Administration) funded research
and development for deploying traditional CPU applications to GPU-based
exascale architectures in United States~\cite{ecp}.

A common strategy in GPU-based science applications is to setup the problem
(a one time cost that can usually be amortized over application runtime)
on the CPUs and execute compute intensive tasks on the GPUs.
Flexibility provided by CPU programming models is essential during problem
setup as it involves creating and initializing data structures used later in
application execution.
In contrast, compute intensive tasks with a sufficient level of parallelism
can benefit from higher memory bandwidth and floating point performance
offered by GPUs.
Explicit synchronization between CPU and GPU memory is required during
an application (such as copying problem data from CPU to GPU and then copying
solution back from GPU to CPU) as they usually have separate memory address spaces.

The low-level GPU programming models employ ``massively parallel''
\textit{single instruction multiple threads} (SIMT) execution model and require
user to specify work for each thread. 
Thus, the thread-based parallelism is built into the programming model itself
(in contrast to CPU programming where instructions are executed using a single
thread unless the user explicitly uses multiple threads).
During execution, the GPU spawns a set of threads (the size is determined by
the problem size) all of which will execute the same instruction on different
data.
The threads are organized into thread blocks each of which having of a fixed
number of threads.
\Cref{fig:vec_scale} shows a simple vector scaling kernel written with C for
CPUs (left) and CUDA~\cite{nvidia_cuda_guide} (right) for GPUs.
The \texttt{for} loop in C is replaced by a single index calculation in CUDA
which involve the thread block index (\texttt{blockIdx.x}),
thread block size (\texttt{blockDim.x}) and thread index in the thread block
(\texttt{threadIdx.x}).
This index is used to assign a single iterate of the \texttt{for} loop to
each GPU thread at runtime.
Then the GPU threads execute the loop iterates in parallel.

\begin{figure*}
\centering
\begin{minipage}{.30\linewidth}
\centering
\begin{minted}[fontsize=\scriptsize,framesep=2mm]{C}
for (int i = 0; i < N; ++i) {
  a[i] = 2 * b[i];
}

\end{minted}
\end{minipage}
\hfill \vline \hfill
\begin{minipage}{.65\linewidth}
\centering
\begin{minted}[fontsize=\scriptsize,framesep=2mm]{CUDA}
__global__ void vec_scale(double *a, double *b, int N) {
  int i = blockIdx.x * blockDim.x + threadIdx.x;
  if (i < N)
    a[i] = 2 * b[i];
}
\end{minted}
\end{minipage}
\caption{\label{fig:vec_scale}A vector scaling program written
for CPUs with C (left) and for GPUs with CUDA (right).}
\end{figure*}

Unlike in the embarrassingly parallel kernel in \Cref{fig:vec_scale}, some
kernels (sum reductions for example) will require explicit synchronization
between GPU threads within a thread block.
The programmer may also be forced to explicitly manage the movement of data
across the GPU memory hierarchy in order to efficiently utilize the GPU memory
bandwidth.
The low-level programming models for GPUs (CUDA, HIP~\cite{amd_hip_guide},
OpenCL~\cite{khronos_opencl_registry}, etc.) expose
these low-level synchronizations and data movements but they require
a more involved development process.
The high-level programming models (OpenMP~\cite{openmp52},
OpenACC~\cite{openacc32}, etc.) on the other hand are
comparatively easier to use but they lack the required low-level control
essential for writing high-performance code.
One downside of both low-level and high-level programming models is the lack of
performance portability since both force the programmers to make decisions about
the execution plan (i.e., how the instructions are mapped to hardware) of a
program when the algorithm is expressed using the respective model.

Thus, there are significant challenges in increasing programmer productivity,
application portability, and performance.
We recognize that this is a complex problem, and there is no one-size-fits-all
solution that achieves all three goals.
However, we believe it is possible to improve programmer productivity without
sacrificing performance or portability.
Learning from the experience of porting spectral/finite element method (SEM/FEM)
applications to GPUs, we were able to identify certain features, if available
on a GPU programming model, would improve programmer productivity.
Our approach is to reduce complexity of the programming model by providing a
method to reuse domain specific optimization patterns.
To this end, we propose \nomp: a framework for building domain specific
compilers for writing high performing GPU applications in C programming
language\footnote{\url{https://github.com/nomp-org}}.
The main contributions of our work are:
\begin{itemize}
\item Develop a framework for creating domain specific compilers for
applications written in the C programming language.
\item Improve programmer productivity by reusing kernel optimization patterns
in a given domain.
\item Achieve comparable performance to optimized programs written in
low-level programming models provided by hardware vendors.
\end{itemize}

\section{Survey of GPU programming models}\label{sec:related}
Compiler pragmas (supported by high-level languages such as C, C++, and
Fortran) are special instructions in the source code intended for the
pre-processor.
Pragma based meta-programming has been used for separating semantics
of a program from its execution schedule and/or
optimizations~\cite{kruse2018user}.
OpenMP~\cite{openmp52}, arguably the most widely supported pragma based
programming model, can be used to program both CPUs and GPUs.
Latest versions of OpenMP has basic code transformations such as loop
unrolling and tiling~\cite{openmp52}.
The authors in~\cite{kruse2018user} propose (with a prototype implementation)
a user-directed composable loop transformations for OpenMP using
transformations based on LLVM/Polly~\cite{grosser2011polly}.
OpenACC~\cite{openacc32} also use pragmas to offload computation to the
GPUs.

X language proposed in~\cite{donadio2006language} for programming CPUs uses
pragmas to name loops or code segments and to specify the transformations.
These pragmas are then translated in two stages.
First, a frontend compiler parses the annotated source code and convert the
pragmas to library calls.
Then the translated code is executed with a file that describes the
optimizations to be performed.
This process generates the final optimized source code, which is then
compiled and run again.

HMPP~\cite{donadio2006language} is another pragma based language that targets
both CPUs and GPUs.
HMPP provides GPU offloading functionality similar to that of OpenMP,
but HMPP compiles ``codelets'' (functions which must be run on the GPUs) to
all the available hardware before the executable is run, thus inhibiting any
chance of performing runtime optimizations.

\section{Reusable Optimization Patterns}\label{sec:patterns}
The kernel optimization patterns are usually reused across different
algorithms in the same domain.
We define a ``kernel'' as an implementation of an algorithm for a given
hardware platform.
This is made possible by the fact that the algorithms in a given domain
relate the same physical quantities derived from a common mathematical
formulation.
For example, SEM/FEM kernels relate quantities such as mesh elements (resulting
from the spatial discretization), the degrees-of-freedom (depends on spatial
discretization and approximation order) and derivative matrices
(which depends on the polynomial order).
Subsequently, the same set of kernel optimizations often work well for
different algorithms in that domain.

In SEM/FEM applications, element-wise kernels (such as mass and stiffness
operators) are parallelized over mesh elements.
Usually, the outer most loop in the kernel iterate over all the
mesh elements in the problem.
We refer to this type of loops as {\em element-loops}.
We also encounter parallelization over the local degrees-of-freedom within each
element and we denote the loops used for this purpose as {\em 1d-dof-loops}.
The number of {\em 1d-dof-loops} in a kernel depends on the physical dimension
of the problem.
In GPU-based SEM solvers (e.g., NekRS~\cite{fischer2021nekrs},
libParanumal~\cite{chalmers2020libparanumal}), the {\em element-loop} is always
parallelized over thread blocks (or group of threads).
In CUDA terminology, this is equivalent to mapping the
{\em element-loop} to the \texttt{blockIdx.x} axis.
{\em 1d-dof-loops} are then parallelized over threads within a thread block.
One option for parallelizing the {\em 1d-dof-loops} is to
map them to \texttt{threadIdx.x}, \texttt{threadIdx.y} and \texttt{threadIdx.z}.
For instance, a problem with $E=8000$ elements using approximation
order $N=7$ (8 degrees-of-freedom in each direction) in 3D
will be parallelized over 8000 thread blocks with each thread block having
$8\times 8\times 8$ threads.

In order to reuse optimizations, first the programming model must have a
mechanism to denote the metadata for domain quantities present in a kernel.
For example, in SEM/FEM kernels, programmers should be able to denote
{\em element-loops} and {\em 1d-dof-loops}.
Then the programming model (or relevant runtime) should be able to transform
the kernel to an optimized version by using metadata to find and apply
relevant optimization patterns in the domain.
Therefore, the programming model should also have a mechanism for users to
specify domain specific optimization patterns.
We also need an intermediate representation (IR) for the kernel that can be
manipulated based on the user provided metadata and optimization patterns.
The role of the IR is to capture the mathematical
representation of the kernel and manipulate this representation
as necessary during the optimization process.
The \nomp~framework uses loopy~\cite{klockner2014loo} IR for code
transformations.

We opted for a pragma-based programming model for the \nomp~framework
primarily due to the popularity and smaller learning curve
of OpenMP~\cite{openmp52,dagum1998openmp}.
In \nomp, we use pragma directive and clauses
to pass kernel metadata as key-value pairs to \nomp~runtime.
Here the key represents the metadata (e.g. {\em element-loop}, etc.) and
the value denotes the variable name associated with the key.
Furthermore, pragmas can be ignored by compilers that do not support them
without generating errors.
Therefore, a program using \nomp~can be compiled with any C compiler without
changes to the source code and/or the build process whether the compiler
supports \nomp~pragmas or not.
This improves the maintainability of the source code.
The two main components of the nomp framework: frontend compiler (\nompcc)
and the runtime (\libnomp) are discussed in \Cref{sec:nomp_nompcc}
and \Cref{sec:nomp_libnomp} respectively.

\section{\nompcc: Clang Based Frontend Compiler}\label{sec:nomp_nompcc}
We modified Clang~\cite{lattner2008llvm} open-source compiler frontend to
recognize \nomp~directives and clauses during its parsing stage.
\nompcc~is this modified Clang compiler frontend that pre-processes
\nomp~pragma directives and clauses, and then compiles the resulting C source.
\nomp~directives and clauses are used to denote C kernels intended for
GPU execution, perform CPU -- GPU data transfers, and provide
metadata about the loops and arrays within the C kernel.
When a nomp directive is detected in the parsing phase, a node(s) for the
corresponding \libnomp~API call(s) is/are inserted into the abstract syntax
tree (AST) of the program.
The nomp directives are prepended with \texttt{\#pragma nomp} 
and must adhere to the syntax listed in \Cref{fig:pragma_syntax}.
The placeholders for constants, variables, expressions, operations and
strings used with the nomp pragmas are denoted within angle brackets
(\texttt{<>}) in \Cref{fig:pragma_syntax}.

\begin{figure}[H]
\centering
\begin{minted}[fontsize=\scriptsize,framesep=2mm]{C}
#pragma nomp init(<argc>, <argv>)

#pragma nomp update(<update_operation>: \
  [<array>(<start_index>, <end_index>)]+)

#pragma nomp for [name(<name>)] \
  [transform(<script_name>, <function_name>)] \
  [annotate(<key>, <value>)] [jit(<variable>)] \
  [reduce(<array>, <operation>)] \

#pragma nomp sync

#pragma nomp finalize
\end{minted}
\caption{Syntax of the nomp directives and clauses supported by nompcc.}
\label{fig:pragma_syntax}
\end{figure}

The \texttt{init} directive initializes the \libnomp~runtime.
It accepts C-style command line arguments containing runtime
configurations (which GPU and/or backend to use, verbosity level etc.).
\nompcc~replaces the \texttt{init} directive by a
call to \texttt{nomp\_init()} \libnomp~function.
The \texttt{update} directive is used to allocate memory on the GPU,
copy data to and from the GPU, and free the allocated memory on the GPU.
The \texttt{update} directive accepts a memory operation, followed by a
list of one or more variables each with a start and an end index.
The operations supported by \texttt{update} is shown in
\Cref{tab:update_op}.
\nompcc~replaces \texttt{update} directive by call(s) to \texttt{nomp\_update()}
function.

\begin{table}[htb]
\centering
\small
\begin{tabular}{ll}
\toprule
\textbf{Operation} & \textbf{Description} \\
\midrule
\texttt{alloc} & Allocate memory on the GPU (no copy). \\
\texttt{to} & Copy data to the GPU. \\
\texttt{from} & Copy data from the GPU. \\
\texttt{free} & Free allocated memory on the GPU.\\
\bottomrule
\end{tabular}
\caption{Memory operations supported by \texttt{update} directive.}
\label{tab:update_op}
\end{table}

The \texttt{for} directive designates a loop nest in the C source
as a GPU kernel.
\nompcc~replaces the \texttt{for} directive by a call to a \texttt{nomp\_jit()}
function followed by a call to a \texttt{nomp\_run()} function.
The loop nest is passed to \texttt{nomp\_jit()} function as a string
along with necessary metadata.
Then the call to \texttt{nomp\_run()} function executes this kernel on the GPU.
The \texttt{for} directive accepts a set of additional clauses listed in
\Cref{tab:for_clauses}.
\texttt{for} directive and clauses only affect a single for-loop nest that
immediately follows them.
The \texttt{sync} directive implements a barrier on the CPU until a GPU kernel
or a non-blocking memory transfer finishes execution.
The \texttt{finalize} directive finalizes \libnomp~runtime and free allocated
resources.
\nompcc~replaces \texttt{sync} and \texttt{finalize} directives by calls
to \texttt{nomp\_sync()} and \texttt{nomp\_finalize()} \libnomp~functions
respectively.
Neither \texttt{sync} nor \texttt{finalize} accepts any arguments.

\begin{table*}\centering
\small
\begin{tabular}{ll}
\toprule
\textbf{Clause} & \textbf{Description} \\
\midrule
\texttt{transform} & Provides the kernel specific transformation script
and function name. \\
\texttt{annotate} & Provides metadata about loops and arrays
in the kernel.\\
\texttt{jit} & Denotes constant variables for the entire
runtime of the program. \\
\texttt{reduce} & Denotes the reduction variable and operation if the kernel
  has a reduction. \\
\texttt{name} & Name the kernel (useful when debugging).\\
\bottomrule
\end{tabular}
\caption{Clauses supported by \texttt{for} directive.}
\label{tab:for_clauses}
\end{table*}

\section{\libnomp: Loopy Based Runtime Library}\label{sec:nomp_libnomp}
\libnomp~is the runtime library which implements the functions of the
user facing \nomp~pragmas and directives.
\libnomp~is responsible for generating GPU kernels, dispatching
kernels to GPUs using low-level programming models (OpenCL, CUDA, and HIP),
managing CPU -- GPU data transfers, and synchronization.
\libnomp~is a general C library independent of \nompcc, and can be used from
any C program, not just those compiled with \nompcc.

Before generating a low-level kernel, \libnomp~converts the C source code
into an intermediate representation (IR) based on loopy~\cite{klockner2014loo}.
During program execution, \libnomp~consults user authored domain and kernel
specific transformations scripts to optimize the IR based on the metadata
passed into \libnomp.
Once the IR is finalized, \libnomp~uses loopy to generate low-level GPU
source.~\libnomp~then compile and execute the source on GPUs using
\libnomp~backends.

\subsection{loopy: Transformation and code generation engine}
\label{subsec:nomp_loopy}
loopy is a code transformation and generation engine designed
for array-based code found in scientific computing
applications such as
dense/sparse linear algebra, convolutions, $n$-body and
PDE solvers~\cite{klockner2014loo}.
The primary goal of loopy is to generate high performant code for a
variety of GPU and CPU hardware~\cite{klockner2014loo}.
To facilitate this optimization process, loopy has the ability to
represent kernels in different, yet mathematically
equivalent, forms through the loopy intermediate representation
(loopy IR).
loopy provides a set of transformations that operates on loopy IR
to convert the kernel into a
more performant version for the target hardware~\cite{klockner2014loo}.
These transformations include (but not limited to) loop tiling,
loop unrolling, data layout transformations, and loop to hardware
axis mapping.

The loop variables are referred to as ``inames'' in loopy.
loopy offers the ability to map inames to hardware axes as part of
its transformations by specifying ``iname implementation tags''
for each iname.
The iname implementation tags roughly correspond to the parallel axes
defined by the low-level GPU programming models (CUDA, OpenCL, etc.).
\Cref{tab:hw_axes} illustrates several iname implementation tags along
with their OpenCL equivalent parallel
axes~\cite{loopy_doc}.

\begin{table*}
\centering
\small
\begin{tabular}{lll}
\toprule
\textbf{Tag} & \textbf{Meaning} & \textbf{OpenCL equivalent}\\
\midrule
for  & Sequential loop & Thread local loop \\
ord  & Order-enforced sequential loop & Thread local loop\\ 
l.N  & $N^{th}$ local (intra-group) axis &
$N^{th}$ thread block (local) axis i.e., \texttt{local\_id(N)}\\
g.N  & $N^{th}$ global (inter-group) axis &
$N^{th}$ grid (global) axis i.e., \texttt{group\_id(N)}\\
\bottomrule
\end{tabular}
\caption{Iname implementation tags in loopy.}
\label{tab:hw_axes}
\end{table*}

The list of instructions in a loopy kernel are represented as assignment
operations between a left hand side (LHS) and a right hand side (RHS).
RHS can consists of standard mathematical operations, common
functions supported by OpenCL and other
user-defined functions~\cite{klockner2014loo}.
In addition to above expressions, loopy supports reductions through
special reduction instructions
such as \texttt{sum} and \texttt{product} (which implement reductions
over addition and multiplication respectively).
After parsing the C kernel passed into \libnomp~with libclang~\cite{libclang},
\libnomp~then emits relevant loopy instructions for each C kernel AST node.
Once the loopy kernel is generated, loopy API can be used to apply
transformations based on domain/kernel specific transformation scripts.
loopy then generates a GPU kernel for the selected \libnomp~backend.

\subsection{Domain and kernel specific transformation scripts}
\label{subsec:nomp_transforms}
There are two ways to specify the loopy transformations (optimization
patterns) in \libnomp: using a domain specific transformation script and/or
kernel specific transformation script.
These scripts are written in Python using the loopy API.
\Cref{fig:domain_transform} shows an example of a domain specific
transformation script for SEM applications
based on the patterns described in \Cref{sec:patterns}.
The script has a single function called \texttt{annotate()} which is
called on all the kernels in the application.

\begin{figure}[htb]
\centering
\begin{minipage}{.90\textwidth}
\begin{minted}[fontsize=\scriptsize,framesep=2mm,linenos]{python}
import loopy as lp

def annotate(kernel, annotations, context):
    inames = kernel.default_entrypoint.all_inames()
    block_size = min(512,
      context["device::max_threads_per_block"])
    axis = 0
    for key in annotations:
        if key == "1d_dof_loop":
            loop = annotations[key]
            if loop in inames:
                lp.split_iname(kernel,
                  loop,
                  block_size,
                  inner_tag=f"l.{axis}")
                axis += 1
        if key == "element_loop":
            loop = annotations[key]
            if loop in inames:
                kernel = lp.tag_inames(kernel,
                  [(loop, "g.0")])
    return kernel
\end{minted}
\end{minipage}
\caption{A domain specific transformation script for the spectral element
applications. The script defines the \texttt{annotate()} function that
is called
on every kernel in the application.}
\label{fig:domain_transform}
\end{figure}

The \texttt{annotate()} function is called with three inputs:
\texttt{kernel}, \texttt{annotations} and \texttt{context}.
\texttt{kernel} is the loopy kernel object generated from the C AST.
\texttt{annotations} is a dictionary that maps patterns in the domain
(such as {\em 1d-dof-loop} and {\em element-loop} to a corresponding 
loop variable (iname) or a data structure (array).
These annotations are used inside the \texttt{annotate()}
function to apply domain specific transformations on the kernel.
For example, in line $20$--$21$ of \Cref{fig:domain_transform},
we map the {\em element-loop} to the global axis $0$.
\texttt{context} is a dictionary of metadata about the backend and the
GPU used to execute the kernel.
This information can be used to apply transformations based on the
GPU parameters such as maximum threads per block supported by the GPU.
\Cref{tab:context_params} summarizes the metadata provided in the
\texttt{context} dictionary.
Finally, the \texttt{annotate()} function returns a modified \texttt{kernel}
object.

\begin{table*}\centering
\small
\begin{tabular}{lp{10cm}}
\toprule
\textbf{Key} & \textbf{Description of the value} \\
\midrule
\texttt{device::max\_threads\_per\_block} & Maximum threads per block supported by the device. \\
\texttt{device::name} & Name of the device. \\
\texttt{device::vendor} & Name of the hardware vendor (e.g., AMD, Intel). \\
\texttt{device::driver} & The device driver version used to run the kernel. \\
\texttt{backend::name} & Name of the backend used to run the kernel (e.g., HIP, etc.). \\
\bottomrule
\end{tabular}
\caption{Parameters in the \texttt{context} dictionary and their description.}
\label{tab:context_params}
\end{table*}

\begin{figure}[htb]
\centering
\begin{minipage}{.90\textwidth}
\begin{minted}[fontsize=\scriptsize,framesep=2mm,linenos]{python}
def gs(kernel, context):
    block_size = min(512,
      context["device::max_threads_per_block"])
    kernel = lp.split_iname(kernel, "i",
      block_size)
    kernel = lp.tag_inames(kernel,
      {"i_outer": "g.0", "i_inner": "l.0",
        "j*": "ord"}
    )
    return kernel
\end{minted}
\end{minipage}
\caption{A kernel specific transformation function for the gather-scatter kernel
used in the Nekbone benchmark implemented in \Cref{sec:nomp_results}.}
\label{fig:kernel_transform}
\end{figure}

\Cref{fig:kernel_transform} shows a kernel specific transformation function
implemented for the gather-scatter kernel (\texttt{gs}) found in the Nekbone
benchmark described in \Cref{sec:nomp_results}.
Unlike the \texttt{annotate()} function, the \texttt{gs()} function does not
require \texttt{annotations} dictionary as an input.
Instead, the function depends on \texttt{gs} kernel specific information to
perform transformations.
The \texttt{gs()} function assumes that the loopy kernel object,
\texttt{kernel}, has an iname named \texttt{i}.
The function splits the iname \texttt{i} into two inames: \texttt{i\_outer}
and \texttt{i\_inner} (line $4$--$5$) with a block size (or tile size)
based on the target hardware (line $2$--$3$).
It then tags the resulting inames with implementation tags \texttt{g.0}
and \texttt{l.0} respectively (line $6$ -- $9$).
The transformation function also expects the kernel to have multiple
inames starting with \texttt{j} and tags all the inames starting
with \texttt{j} with \texttt{ord} tag.

\subsection{Backends: Low-level dispatch to hardware}
\label{subsec:nomp_backends}
\libnomp~uses its backend implementations to compile and execute the
optimized loopy kernel on the target hardware.
These backend implementations are thin wrappers around the low-level APIs
exposed by the hardware vendors and support operations such as device and
memory management, kernel dispatch, and device synchronization.
Currently, \libnomp~implements wrappers for CUDA, OpenCL and
HIP driver and runtime APIs enabling executions of
the kernels on NVIDIA, AMD and Intel hardware.
Thus, the users of the \nomp~framework do not need to modify their application
in order to execute it on different vendor hardware.

The specific details how each backend wraps the respective vendor
drivers and runtimes vary due to the differences in the low-level APIs
provided by the vendors.
CUDA and HIP are similar in their low-level APIs, allowing
libnomp to reuse common functionality across their backend
implementations.
We implemented the core functionality of CUDA and HIP in a header file
using place holder macros for backend functions and types.
Then we include this header file in each CUDA and HIP backend
source file after defining the macros appropriately for each backend.
OpenCL on the other hand has a siginificantly different API for device
and memory management making its backend implementation different from
both CUDA and HIP.

\section{\nomp~workflow}\label{sec:workflow}
\begin{figure}\centering
\includegraphics[scale=0.37]{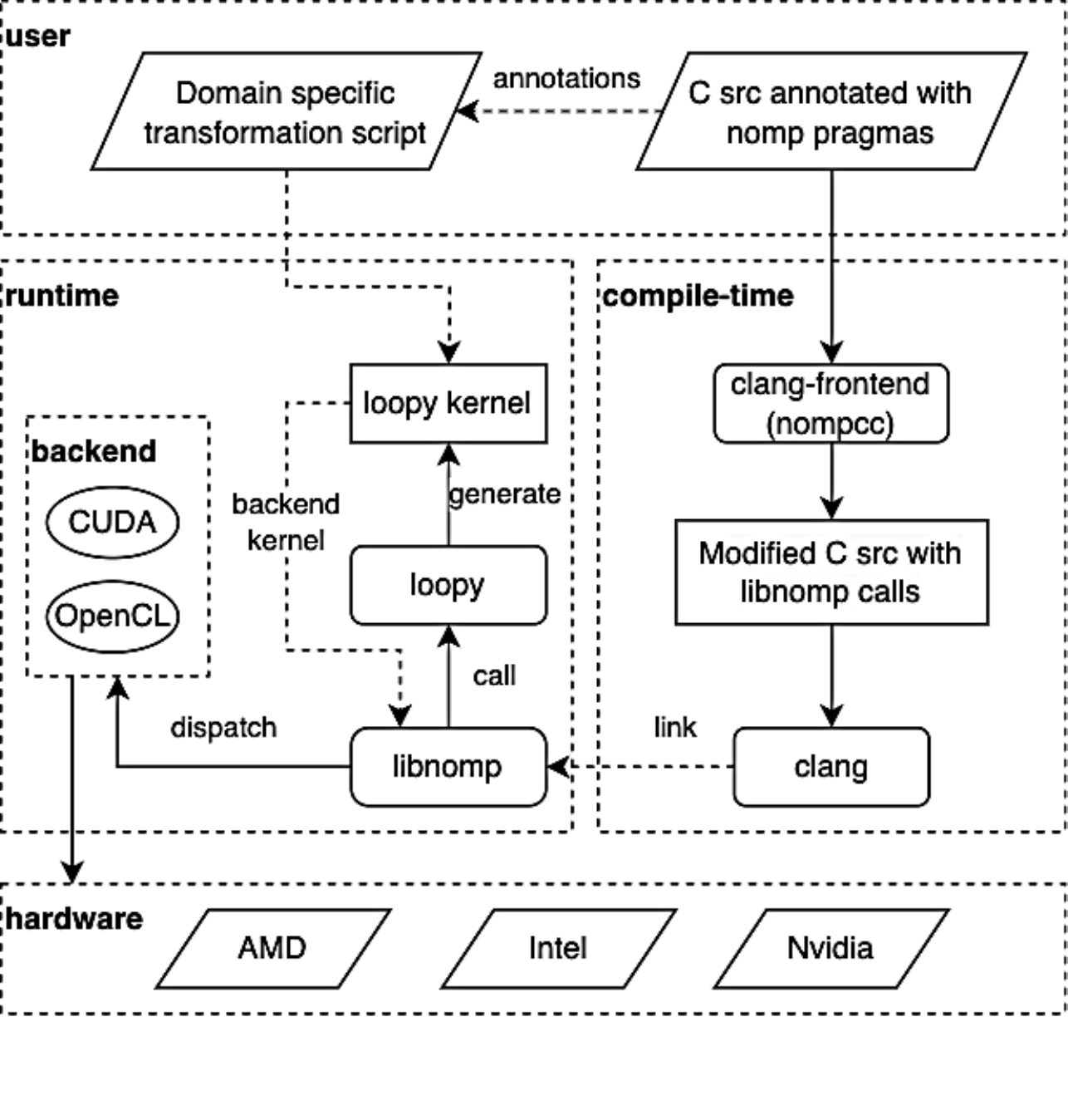}
\caption{\label{fig:nomp_diagram}
\nomp~workflow diagram illustrating user annotating source code,
  compilation with \nompcc, code transformation and generation with loopy
and executing the kernel on hardware using backends.}
\end{figure}

\Cref{fig:nomp_diagram} illustrates the \nomp~workflow, beginning with
the user annotating the C source with \nomp~directives and clauses
followed by compilation of the C program using \nompcc.
\nompcc~links \libnomp~library during the linking phase of the user program.
At runtime, \libnomp~uses loopy for transforming
and generating low-level kernels.
At this phase, loopy consults domain and kernel specific transformation
kernels to generate optimized kernels.
Finally, at runtme \libnomp~compile and execute the generated kernel
on the target hardware using its backend implementations.

\section{Mandelbrot Set Calculation With nomp}\label{sec:nomp_example}
\Cref{fig:mandelbrot} shows an example of calculating the
Mandelbrot set using the \nomp~framework.
The figure illustrates the use of \nomp~pragmas to port computations to the GPU
with minimal modifications to the original C source code.
The \texttt{init} directive on line $2$ initializes the \libnomp~runtime.
The \texttt{update} directive on line $10$ allocates memory on
the GPU for storing the output array \texttt{out}.
Then, on line $12$, the \texttt{for} directive is used to denote the
for-loop nest defined within lines $13$ -- $28$ as a GPU kernel.
The \texttt{for} directive is followed by the \texttt{transform} clause
which specifies the function \texttt{mandelbrot} in the script
\texttt{transform.py} as the kernel specific transformation function.
This \texttt{mandelbrot} function is shown to the right
of \Cref{fig:mandelbrot},
and contains the loopy transformations required to map the C loop nest
to GPU hardware axes.
The \texttt{sync} directive on line $29$ implements a
barrier on the CPU until the GPU kernel completes its execution.
During the execution of the kernel, \texttt{out} array is updated
on line $26$.
The updated \texttt{out} array is then copied back to the CPU using the
\texttt{update} directive on line $30$.
Then, the \texttt{update} directive on line $31$ free the GPU memory
allocated for the \texttt{out} array.
The \texttt{finalize} directive on line $32$ finalizes the \libnomp~runtime.

\begin{figure*}
\centering
\begin{minipage}{.55\textwidth}
  \centering
  \begin{minted}[linenos,fontsize=\scriptsize,framesep=2mm]{c}
void main(int argc, char *argv[]) {
#pragma nomp init(argc, argv)
  float xm = -2, xM = 0.47, ym = -1.12, yM = 1.12;
  int iter = 1000, xres = 1024;
  int yres = xres * (yM - ym) / (xM - xm);
  float dx = (xM - xm) / xres, dy = (yM - ym) / yres;

  int N = yres * xres;
  int *out = malloc(N * sizeof(int));
#pragma nomp update(alloc : out[0, N])

#pragma nomp for transform("transform", "mandelbrot")
  for (int j = 0; j < yres; j++) {
    for (int i = 0; i < xres; i++) {
      float c_im = ym + j * dy, c_re = xm + i * dx;
      float z_re = c_re, t1;
      float z_im = c_im, t2;
      int k = 0;
      for (int m = 0; m < iter; m++) {
        t1 = z_re*z_re - z_im*z_im + c_re;
        t2 = 2 * z_re * z_im + c_im;
        z_re = t1, z_im = t2;
        if (z_re * z_re + z_im * z_im > 4)
          k = m, break;
      }
      out[xres * j + i] = k;
    }
  }
#pragma nomp sync
#pragma nomp update(from : out[0, N])
#pragma nomp update(free : out[0, N])
#pragma nomp finalize
  free(out);
}
\end{minted}
\end{minipage}
\vline
\begin{minipage}{.30\textwidth}
  \centering
  \begin{minted}[fontsize=\scriptsize,framesep=2mm]{python}
  import loopy as lp
  
  LOOPY_LANG_VERSION = (2018, 2)
  
  def mandelbrot(knl):
      knl = lp.split_iname(knl,
          "i", 32,
          outer_tag="g.0",
          inner_tag="l.0")
      knl = lp.split_iname(knl,
          "j", 32,
          outer_tag="g.1",
          inner_tag="l.1")
      return knl
  \end{minted}
\end{minipage}
\caption{Mandelbrot C implementation on GPUs with \nomp~(left)
and loopy-based kernel specific transformations script
\texttt{transform.py} (right).}
\label{fig:mandelbrot}
\end{figure*}

\section{nomp Performance On Nekbone}\label{sec:nomp_results}
We present the performance of \nomp~implementation of Nekbone: a simplified
proxy application (miniapp)~\cite{chalmers2023hipbone,gong2016nekbone}
for the pressure Poisson solve in Nek5000~\cite{nek5000}.
We compare \nomp~performance with native HIP implementations on the
Frontier~\cite{frontier} supercomputer at Oak Ridge Leadership Computing
Facility (OLCF).

\subsection{Nekbone miniapp}\label{subsec:nomp_nekbone}
Nekbone uses a Jacobi (diagonally) preconditioned conjugate gradient (PCG)
method for solving a 3D Poisson equation.
The original Nekbone miniapp was developed with Fortran 77 and
parallelized across multiple CPUs using MPI~\cite{gls99}
following a similar approach to Nek5000.
Physical domain for Nekbone is a 3D box domain discretized into $E$ hexahedral
elements.
Each element is further discretized into $(N + 1)^3$ degrees-of-freedom (DOFs)
using order $N$ polynomial basis functions in each of the three
directions.
The total number of unique DOFs in the domain is approximately $EN^3$ (after
accounting for the duplicated DOFs).
During the solution process, PCG is run for a fixed number of iterations
(100 by default).
We developed a C99 version of Nekbone to compare \nomp~with other low-level
GPU programming models.
Our implementation does not include the Jacobi preconditioner for
the conjugate gradient (CG) method and is exclusively designed
for a single GPU.
Our primary goal is to expose single GPU overheads of \nomp~framework if
there are any.
The Nekbone benchmark repository is open source and accessible on GitHub at
nomp-org/benchmarks\footnote{\url{https://github.com/nomp-org/benchmarks}}.

\subsection{Experimental setup}\label{subsec:nomp_nekbone_setup}
We conducted our experiments on a single GPU compute die (GCD) of an
AMD MI250X GPU on Frontier using \nomp~and HIP based Nekbone
implementations.
We did a comprehensive study by varying the number of elements $E$
and the polynomial order $N$.
For each polynomial order $N$, we varied the number of elements
$E$ from $2^1$ to $2^{14}$.
For a fixed $N$ and $E$, we did 100 iterations of conjugate gradient
(CG) method as a warm up run.
Then we did another 100 iterations of CG and measured the execution time.
We discarded the warm up times as they involve setup and other
library initialization costs.
In addition to the time, we recorded the final residual of the CG method
after 100 iterations.
Based on these experiments, we generate three plots for each polynomial
order $N$.

First we plot residual after 100 iterations of the CG method versus the
number of elements $E$ as a measure of the accuracy of our solver
implementation.
Second plot is the giga ($10^9$) degrees-of-freedom per second (GDOFS)
versus the number of DOFs.
GDOFS is calculated as:
\begin{eqnarray}
\text{GDOFS}=
\frac{EN^3 \times \text{iterations}}{\text{time} \times 10^9}
=\frac{EN^3 \times 100}{t \times 10^{9}}
\label{eq:gdofs}
\end{eqnarray}
Here, $t$ is the time taken for 100 CG iterations.
We calculated $t$ as the average of 5 different runs on Frontier.
Equation~\ref{eq:gdofs} measures the rate of work and is used to compare
performance of each implementation.
Finally, we plot the speedup (or relative performance) of \nomp~over
the native HIP versus the number of elements $E$.
The speedup is calculated as the ratio of GDOFS of \nomp~to GDOFS of HIP
for a given $E$ and $N$ value.

We compared the \nomp~Nekbone implementation with two different versions of
Nekbone implemented with HIP to ensure a fair comparison with the \nomp
implementation.
Since \nomp~uses loopy for runtime code generation,
it can generate more optimized code by fixing runtime parameter values for
loop bounds.
In contrast, HIP requires user intervention to leverage runtime values, either
by using the runtime compilation library provided by HIP (known as
HIP RTC~\cite{hip_rtc}) or by hard-coding the parameter values in kernels at
compile time.
Therefore, we implemented one HIP version with fixed parameter values
(denoted by HIP-fixed) and one without fixed parameter values (denoted by
HIP-variable).

\begin{table*}
\centering
\begin{tabular}{p{10cm}ccc}
\toprule
\textbf{Optimization} & \textbf{nomp} & \textbf{HIP-variable} & \textbf{HIP-fixed}\\
\midrule
Fixed the local group size to $512$ in streaming kernels & yes & no & yes\\
Unrolled local reduction loop in the reduction kernel & yes & no & yes\\
Fixed the polynomial order in Poisson operator kernel & yes & no & yes\\
\bottomrule
\end{tabular}
\caption{Difference between HIP-variable implementation and
HIP-fixed implementation.}
\label{tab:hip_vs_hip}
\end{table*}

\Cref{tab:hip_vs_hip} highlights the differences between HIP-variable and
HIP-fixed implementations.
In \nomp~implementation, local group size is determined at runtime based
on the context information passed into the kernel transformation script
(for example, see line $2$ in \Cref{fig:kernel_transform}).
This fixed value enables runtime kernel optimizations in the
compiler.
Furthermore, since loopy is aware of the local group size at
runtime, it can unroll the tree reduction code within a local
group.
The tree reduction is used in Nekbone when calculating the dot products
used in CG.
We used an unrolled reduction loop similar to \nomp~in the
HIP-fixed implementation.

The results with HIP-fixed and HIP-variable implementations
are presented in \Cref{subsec:nomp_nekbone_results} and
\Cref{subsec:nomp_nekbone_results2} respectively.
The most expensive kernel in Nekbone is the one that performs the action
of the Poisson operator referred to as the \texttt{ax} kernel.
The polynomial order $N$ appears in the \texttt{ax} kernel as the upper
bound for {\em 1d-dof-loops}.
We used the \texttt{jit} clause in
\texttt{nomp for} pragma to inform the \libnomp~runtime that
the value of $N$ should be fixed in the generated code.
Fixing the polynomial order was done just before the code
generation step and after applying all the loopy
transformations to the kernel.
We applied the same optimization in HIP-fixed by hardcoding
the polynomial order during the compile time.

\subsection{nomp vs. HIP-variable}\label{subsec:nomp_nekbone_results}
\begin{figure*}
\centering
\subfloat[Final residual]{{\includegraphics[width=\figratio\textwidth]{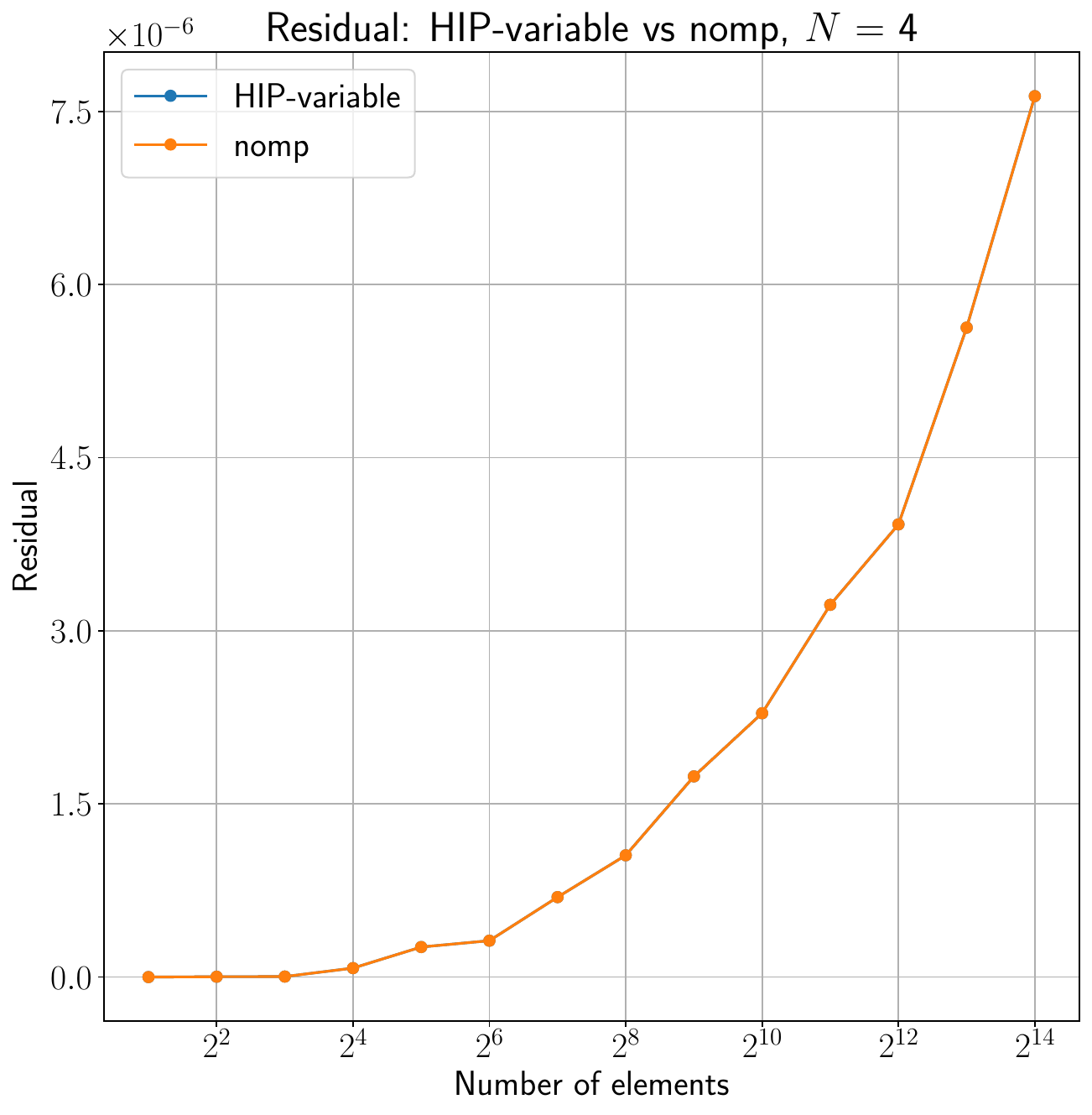}}}
\subfloat[Performance in GDOFS]{{\includegraphics[width=\figratio\textwidth]{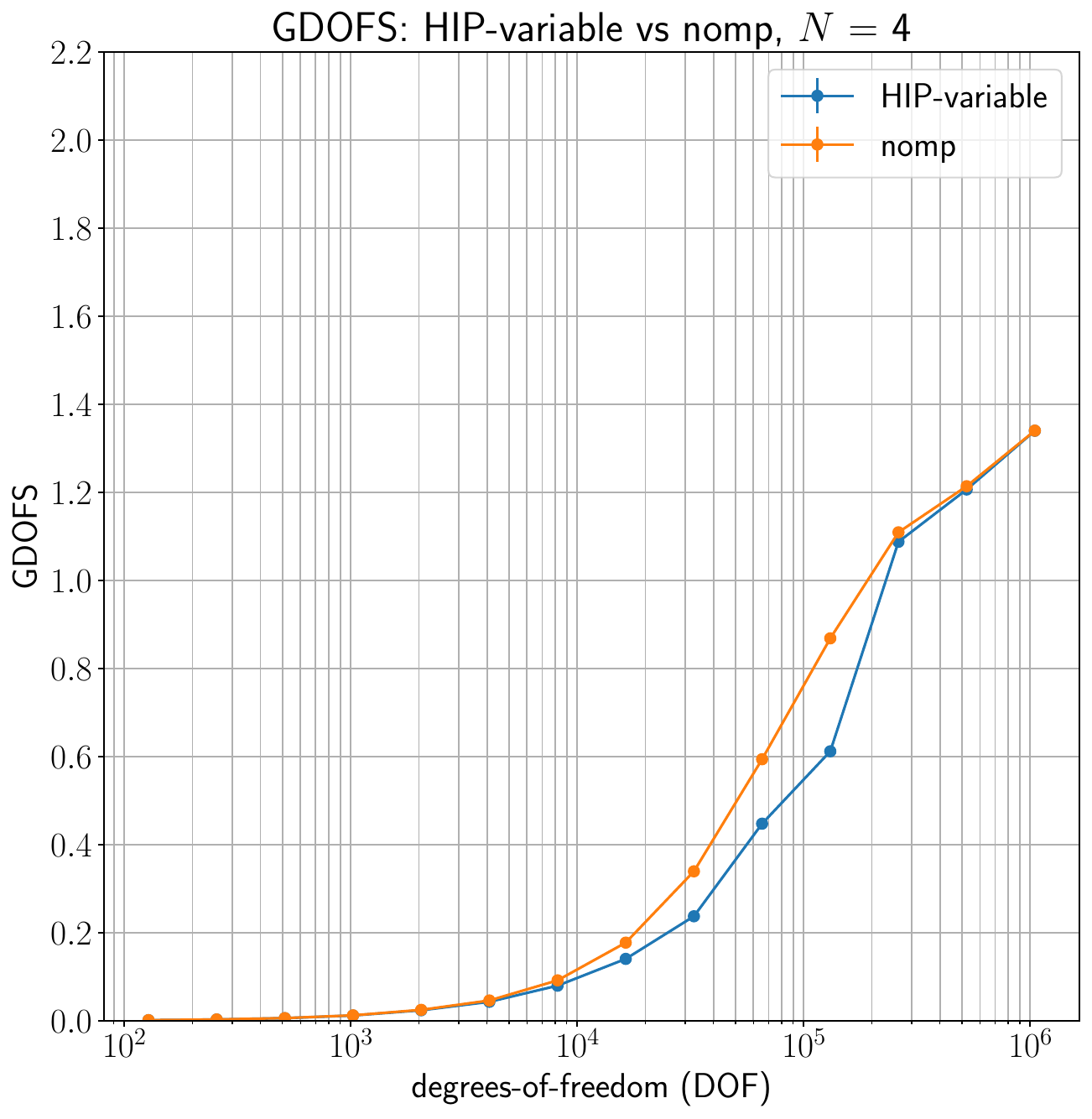}}}
\subfloat[Speedup]{{\includegraphics[width=\figratio\textwidth]{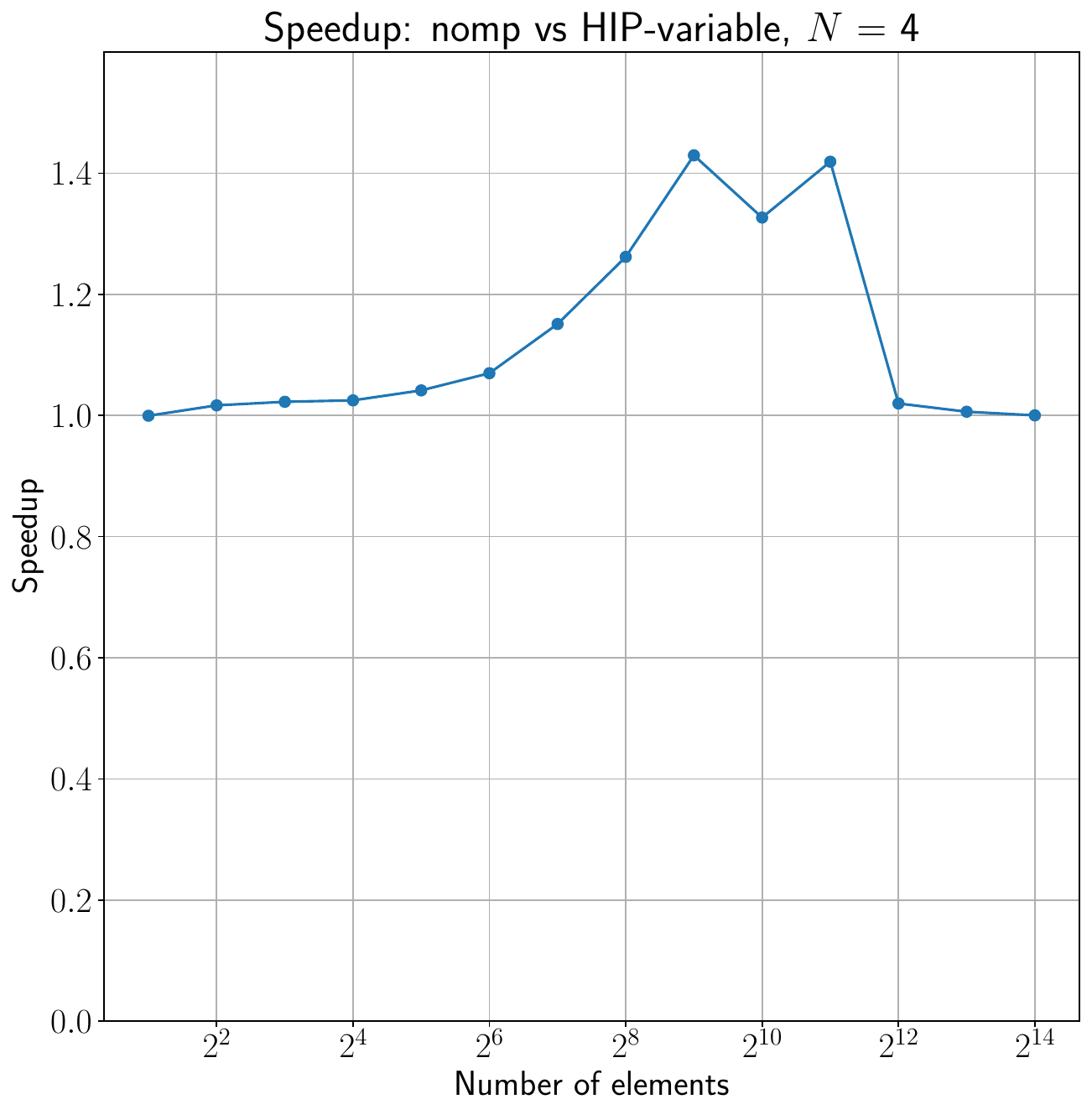}}}
\caption{\label{fig:nomp_N_4} HIP-variable vs. \nomp~Nekbone performance for $N=4$ on a
single GCD of AMD MI250X GPU on Frontier supercomputer at OLCF.}
\end{figure*}

\begin{figure*}
\centering
\subfloat[Final residual]{{\includegraphics[width=\figratio\textwidth]{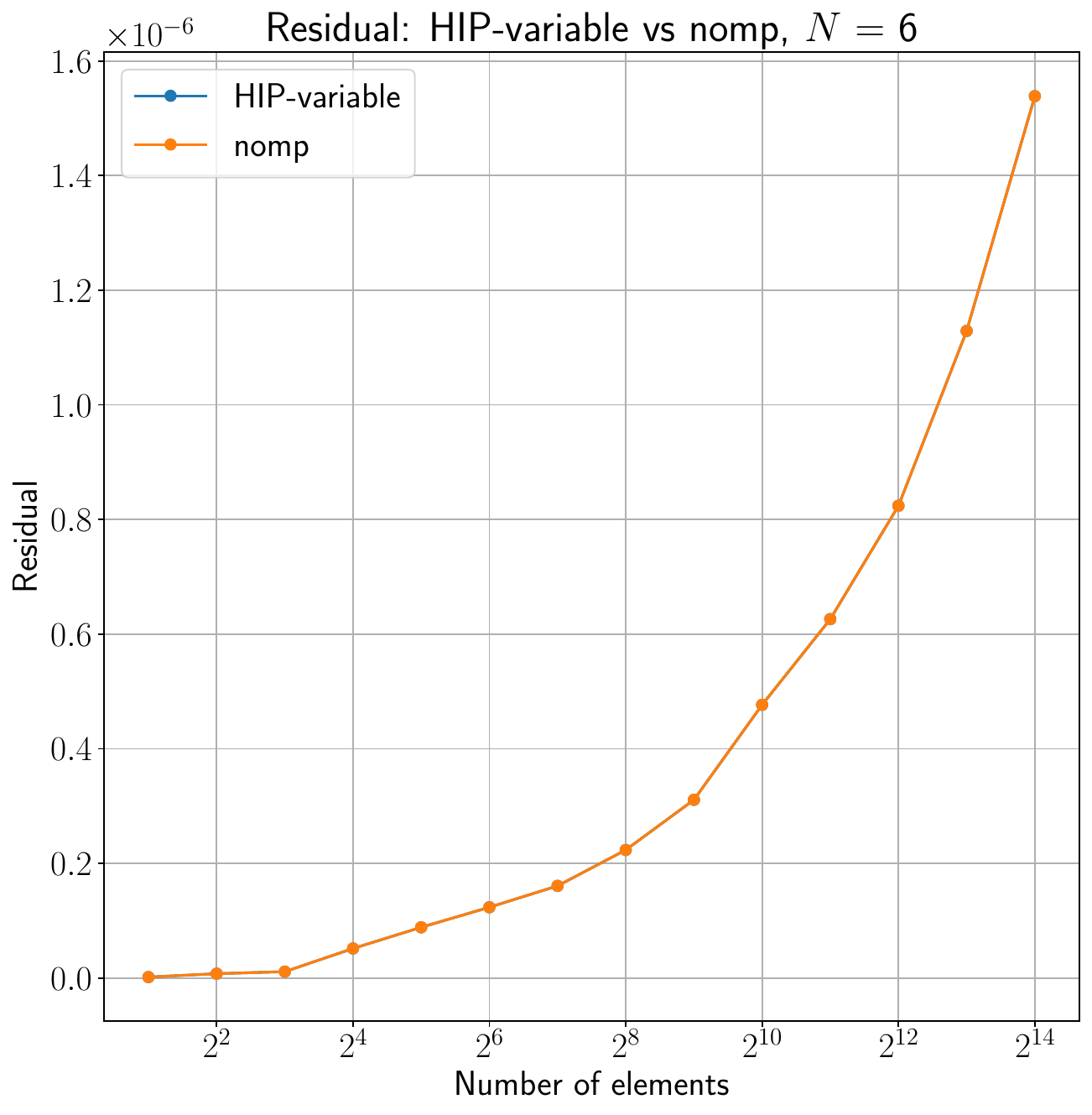}}}
\subfloat[Performance in GDOFS]{{\includegraphics[width=\figratio\textwidth]{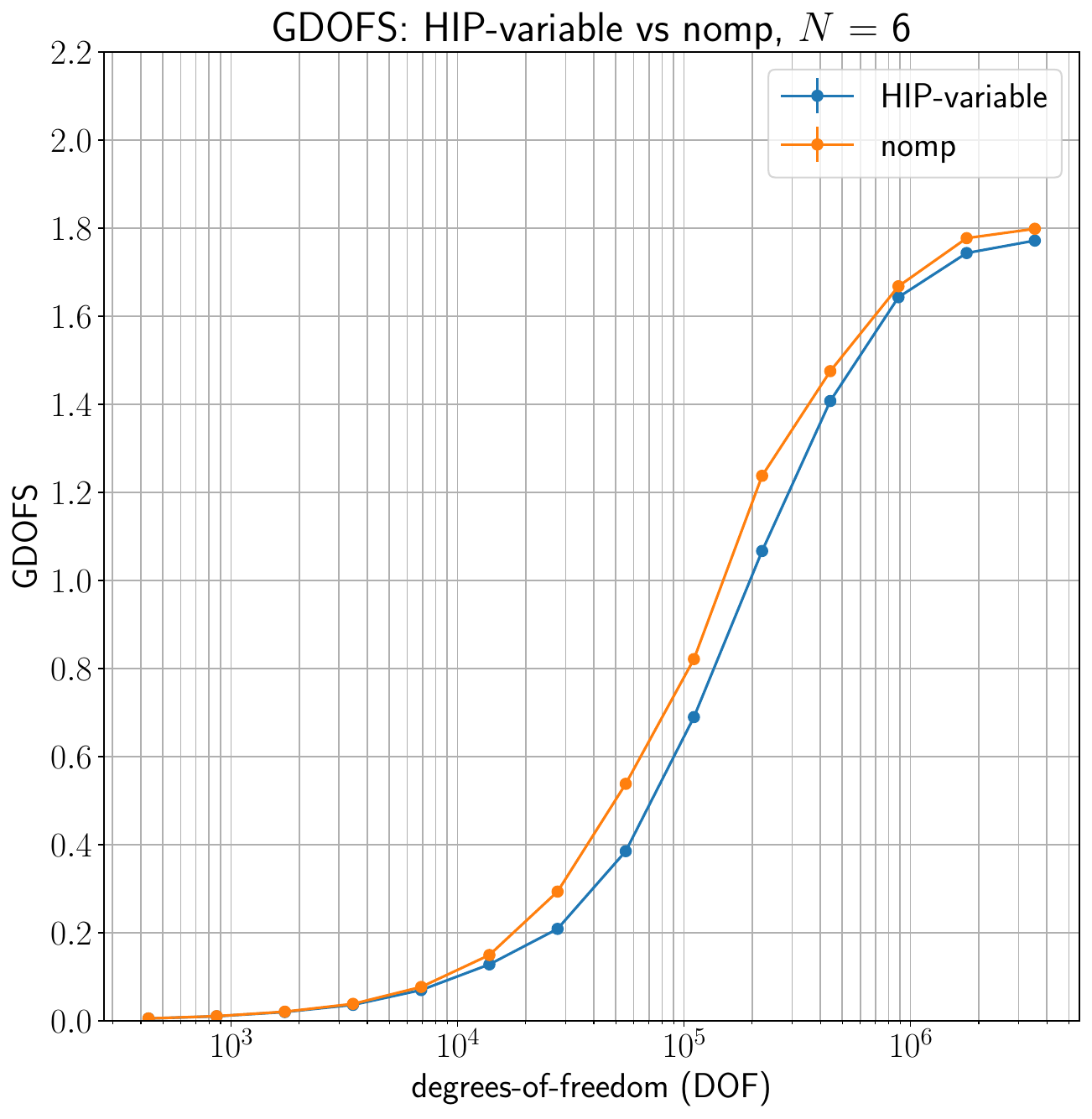}}}
\subfloat[Speedup]{{\includegraphics[width=\figratio\textwidth]{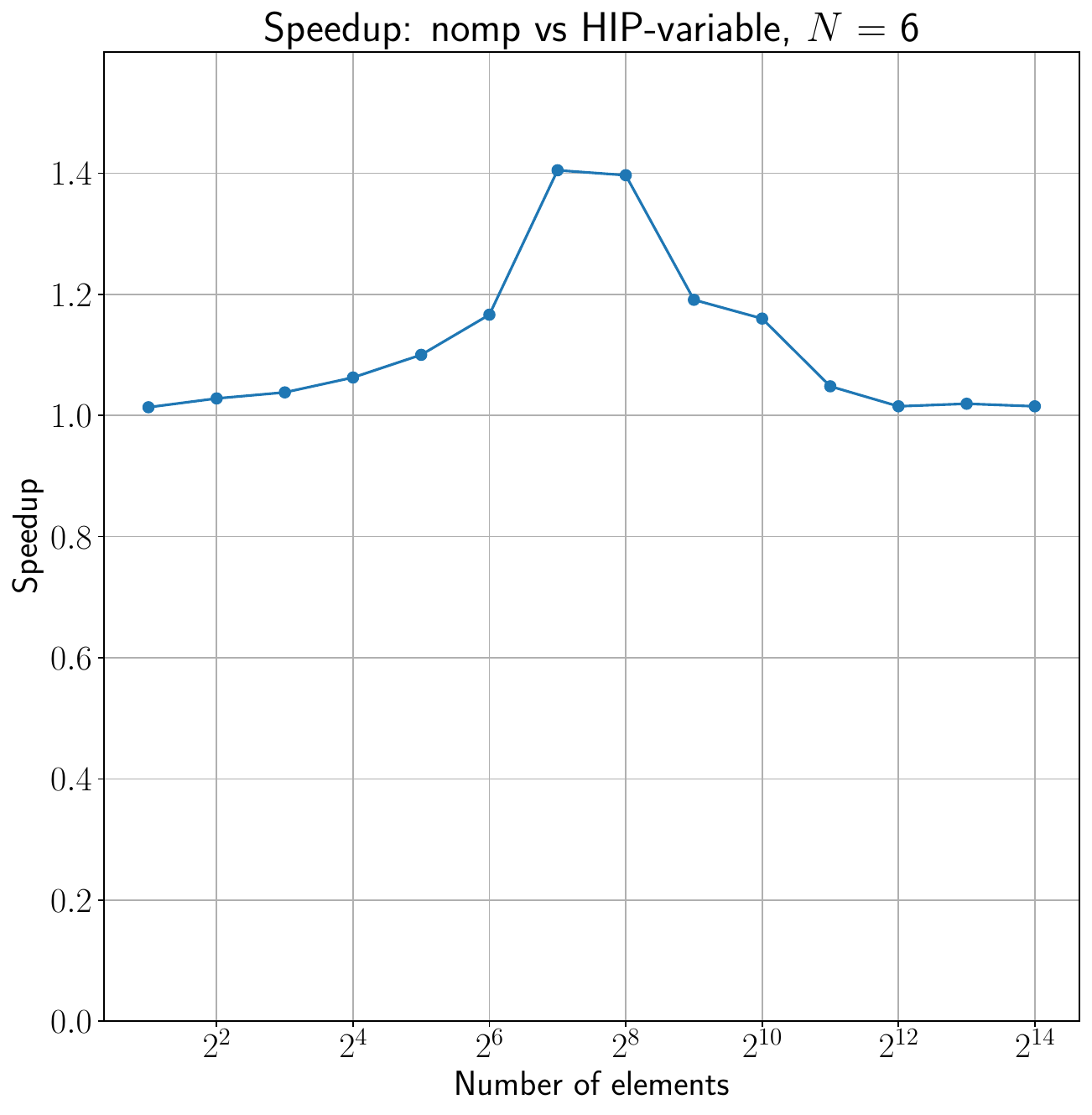}}}
\caption{\label{fig:nomp_N_6} HIP-variable vs. \nomp~Nekbone performance for $N=6$ on a
single GCD of AMD MI250X GPU on Frontier supercomputer at OLCF.}
\end{figure*}

\begin{figure*}
\centering
\subfloat[Final residual]{{\includegraphics[width=\figratio\textwidth]{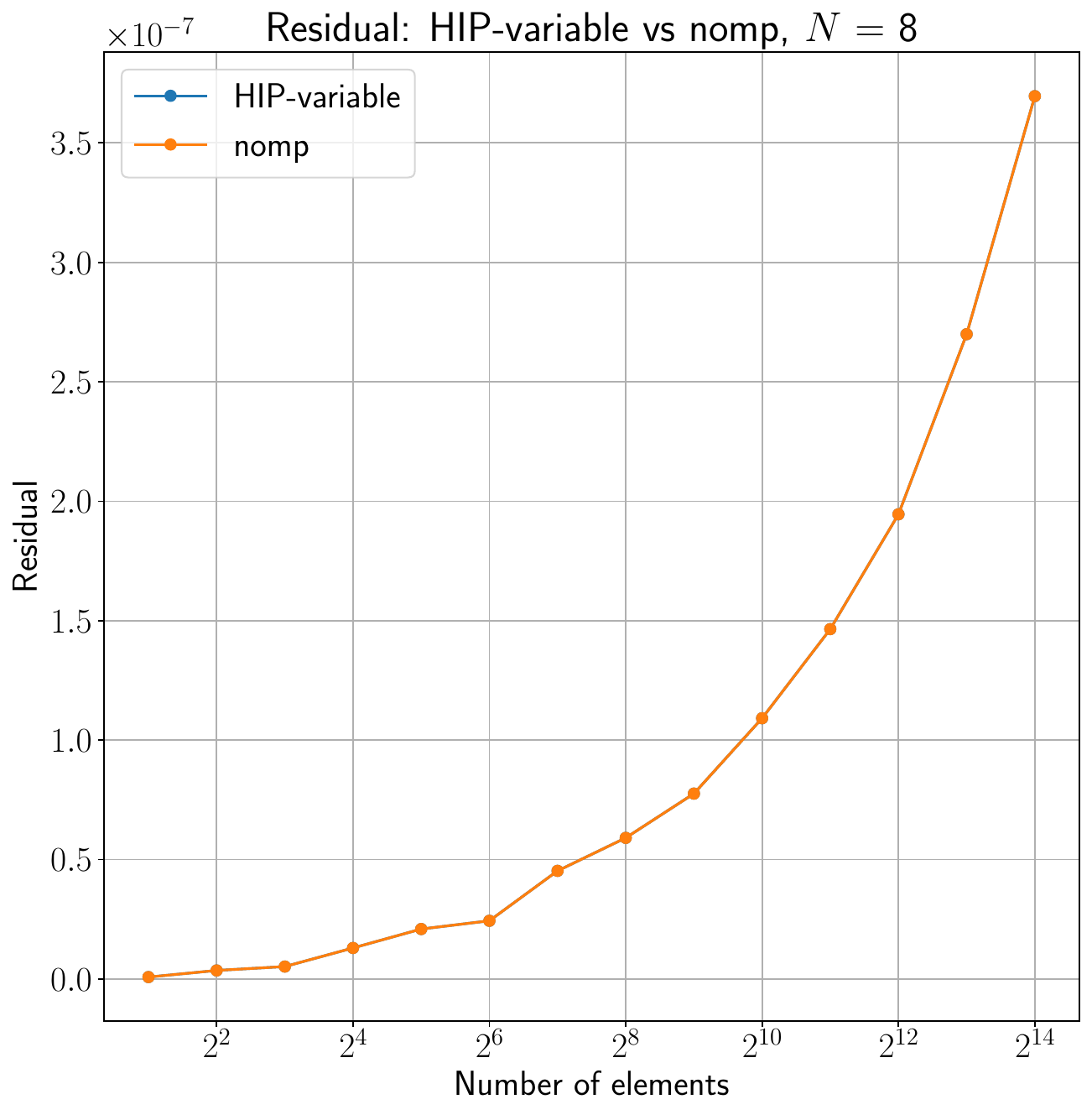}}}
\subfloat[Performance in GDOFS]{{\includegraphics[width=\figratio\textwidth]{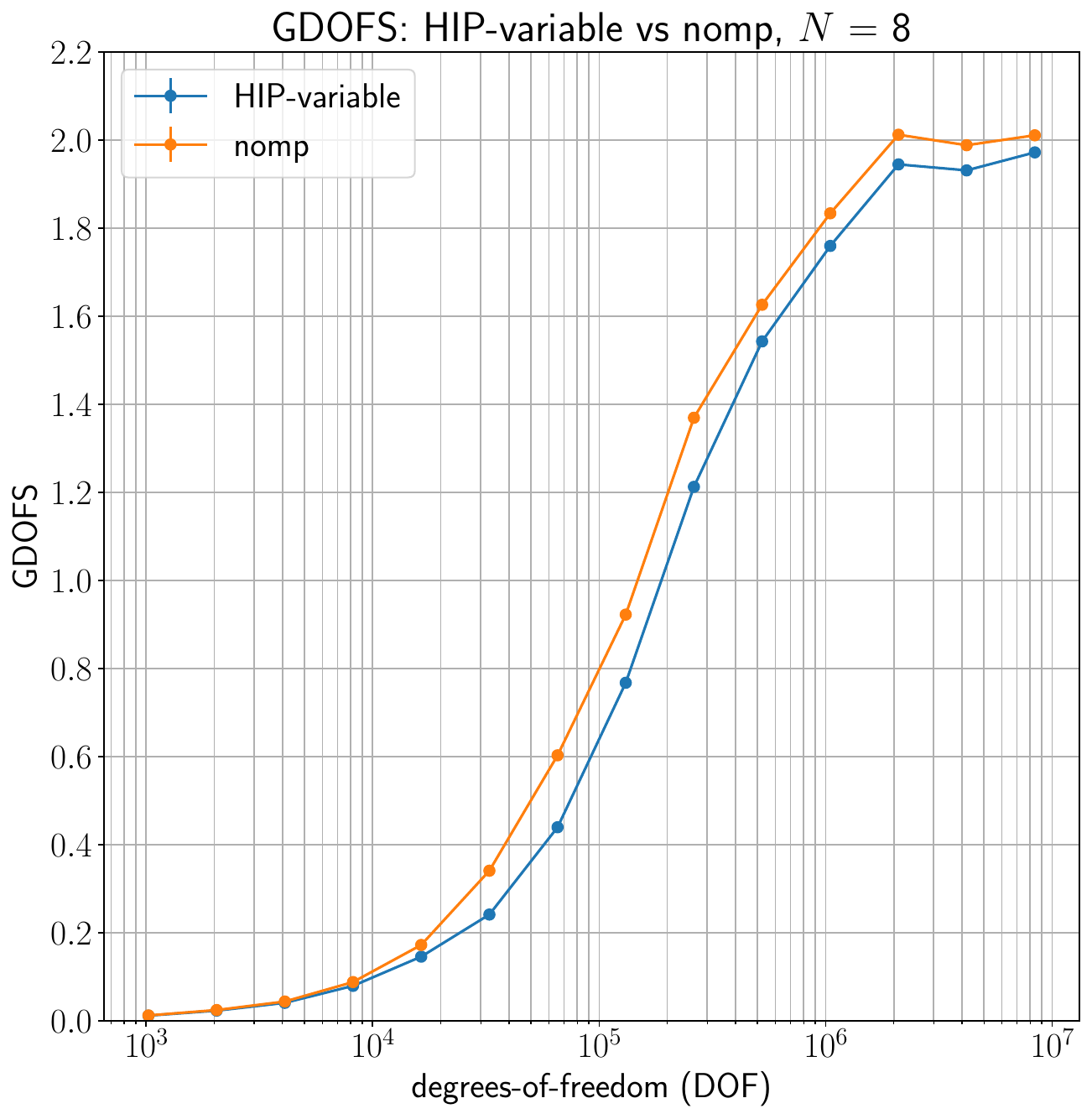}}}
\subfloat[Speedup]{{\includegraphics[width=\figratio\textwidth]{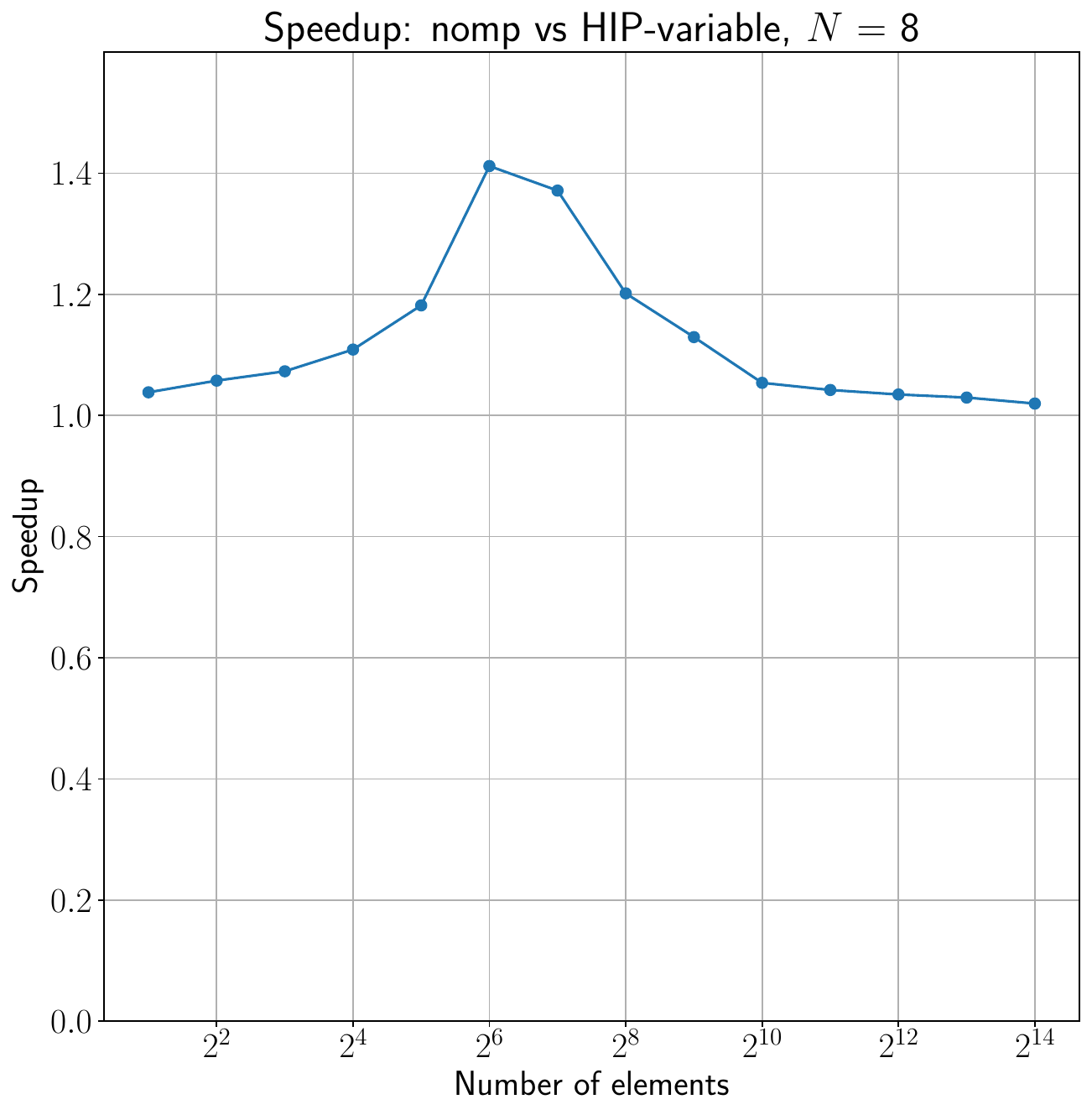}}}
\caption{\label{fig:nomp_N_8} HIP-variable vs. \nomp~Nekbone performance for $N=8$ on a
single GCD of AMD MI250X GPU on Frontier supercomputer at OLCF.}
\end{figure*}

\Cref{fig:nomp_N_4} -- \Cref{fig:nomp_N_8} show \nomp~and HIP-variable
comparison for polynomial orders $N=4, 6$ and $8$.
\Cref{fig:nomp_N_4}(a) -- \Cref{fig:nomp_N_8}(a) indicates that the
residuals are consistent between HIP-variable and \nomp~implementations
across polynomial orders and the element counts.
In addition, the residuals themselves are also within the tolerance of
$10^{-6}$ -- $10^{-7}$ for all of the tests.

\Cref{fig:nomp_N_4}(b) -- \Cref{fig:nomp_N_8}(b) show the GDOFS
versus the DOFs.
The nomp and HIP-variable curves have the same shape, and both curves
show a rapid increase in performance as the DOFs in the
domain increase.
For the large polynomial orders ($N \ge 6$), the performance of both
HIP-variable and \nomp~curves saturate at around $4\times10^6$
DOFs.
For $N=4$, the local problem size (measured in DOFs)
does not reach a sufficiently large value to see performance saturation.
Performance saturates when the problem size is large enough to
reach peak GPU memory bandwidth.
Performance keep increasing with the problem size till the peak bandwidth
is reached.

\Cref{fig:nomp_N_4}(c) -- \Cref{fig:nomp_N_8}(c) show the speedup
of nomp over HIP-variable implementation.
We can see that \nomp~is consistently faster than the HIP-variable
implementation.
This performance gain can be attributed to the optimizations
listed in \Cref{tab:hip_vs_hip} which are naturally applied in
the \nomp~implementation due to runtime code generation
capabilities of loopy.

\subsection{\nomp~vs. HIP-fixed}\label{subsec:nomp_nekbone_results2}
\begin{figure*}
\centering
\subfloat[Final residual]{{\includegraphics[width=\figratio\textwidth]{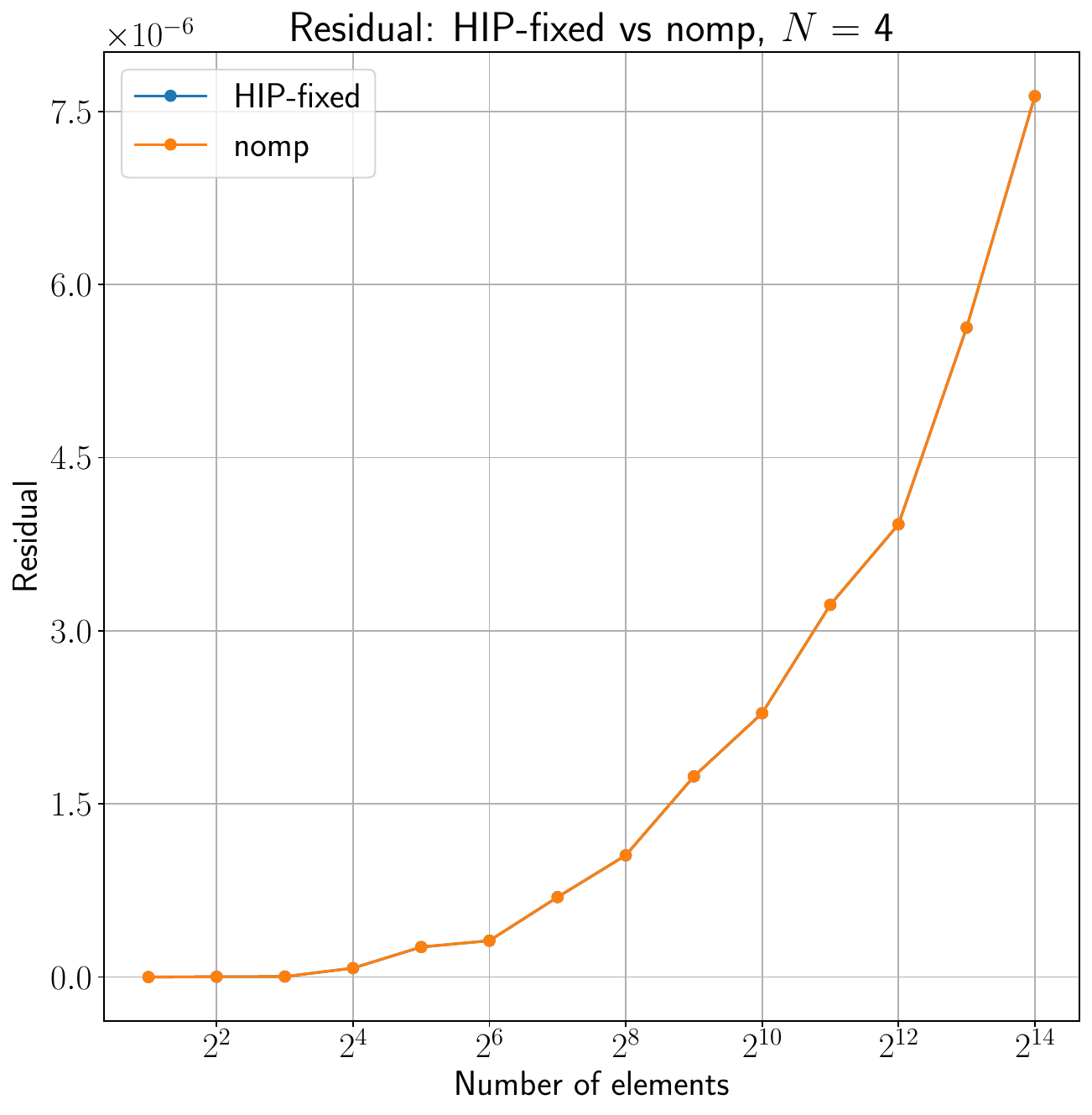}}}
\subfloat[Performance in GDOFS]{{\includegraphics[width=\figratio\textwidth]{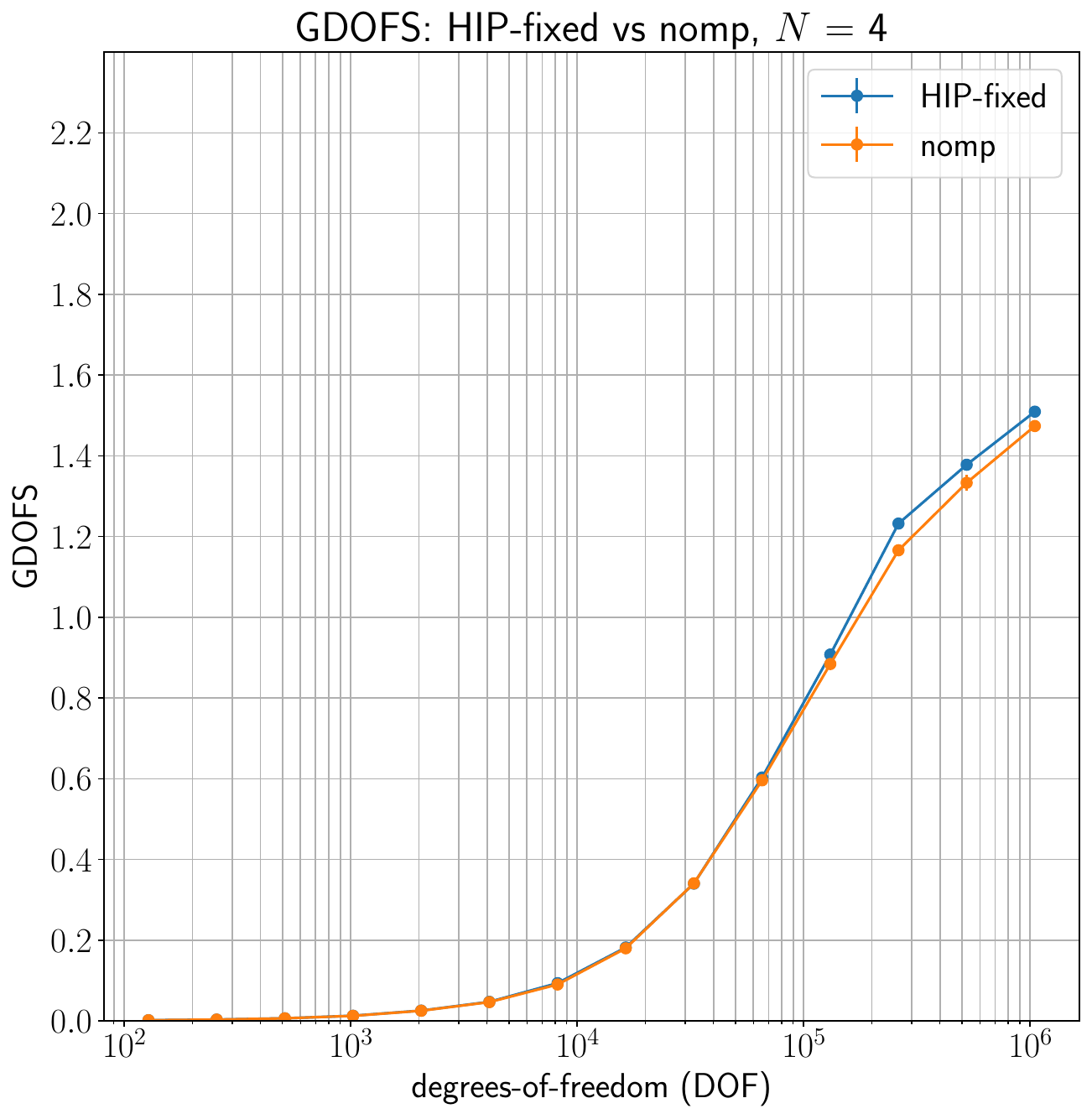}}}
\subfloat[Speedup]{{\includegraphics[width=\figratio\textwidth]{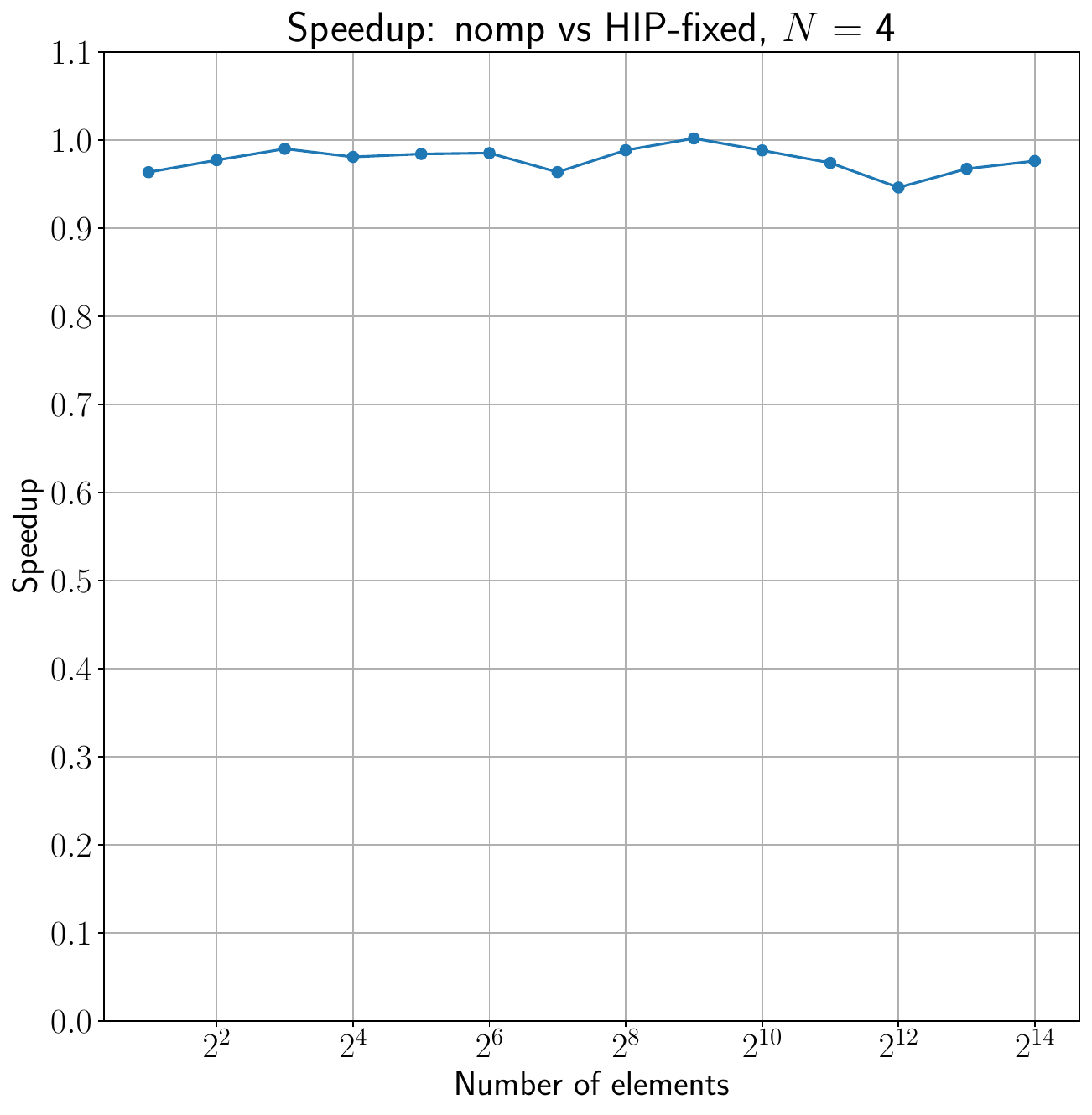}}}
\caption{\label{fig:nomp2_N_4} HIP-fixed vs. \nomp~Nekbone performance for $N=4$ on a
single GCD of AMD MI250X GPU on Frontier supercomputer at OLCF.}
\end{figure*}

\begin{figure*}
\centering
\subfloat[Final residual]{{\includegraphics[width=\figratio\textwidth]{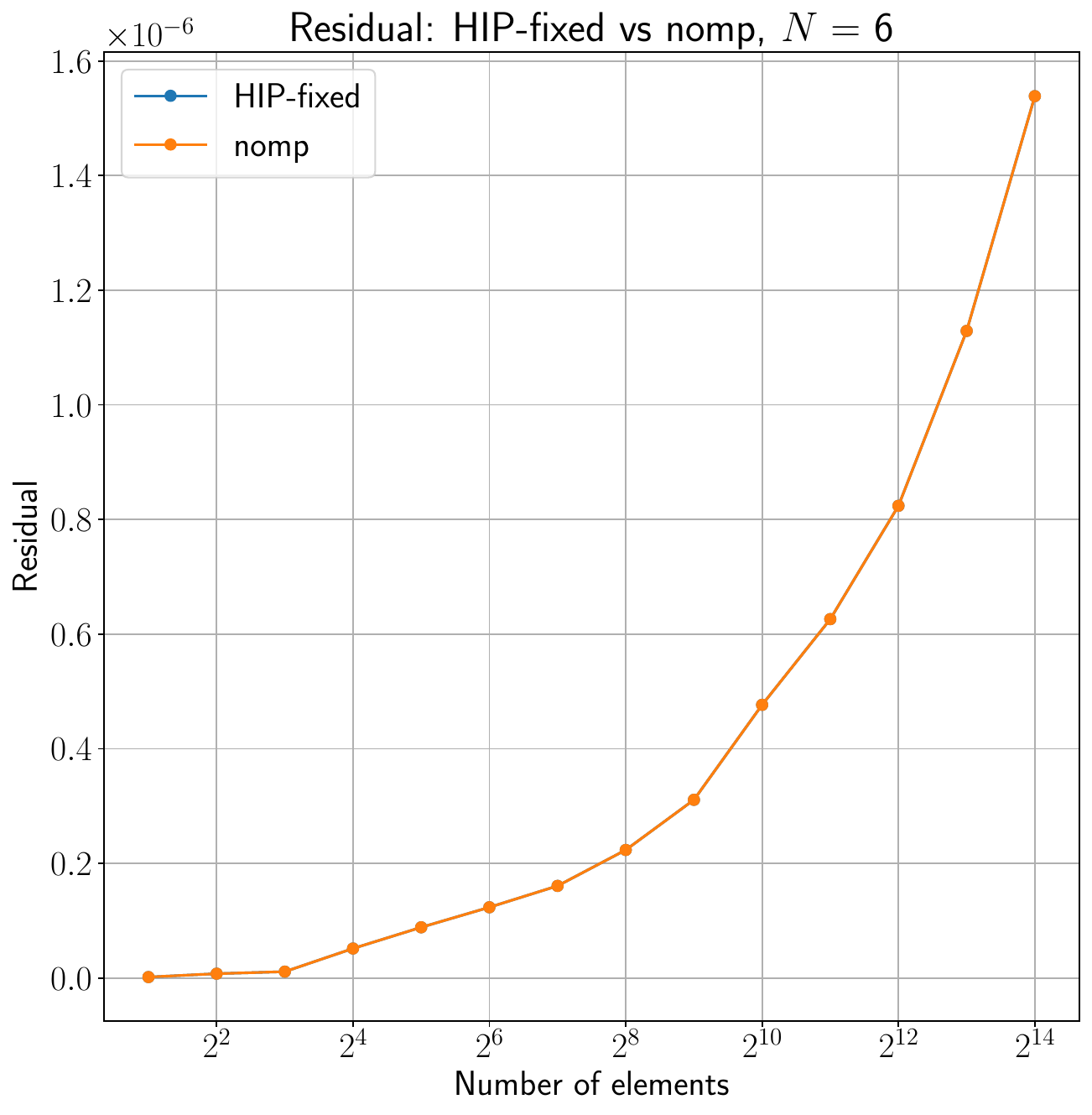}}}
\subfloat[Performance in GDOFS]{{\includegraphics[width=\figratio\textwidth]{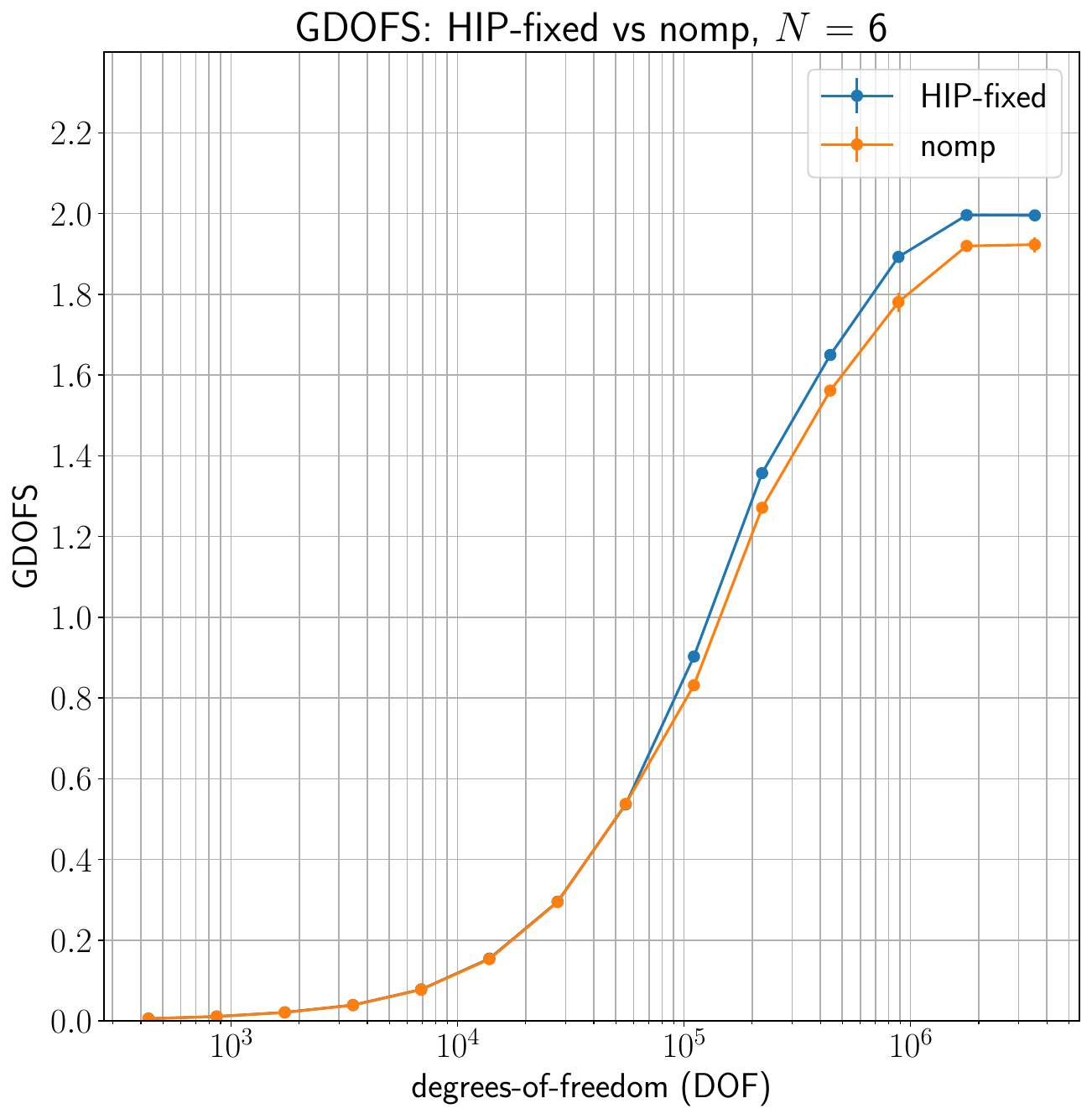}}}
\subfloat[Speedup]{{\includegraphics[width=\figratio\textwidth]{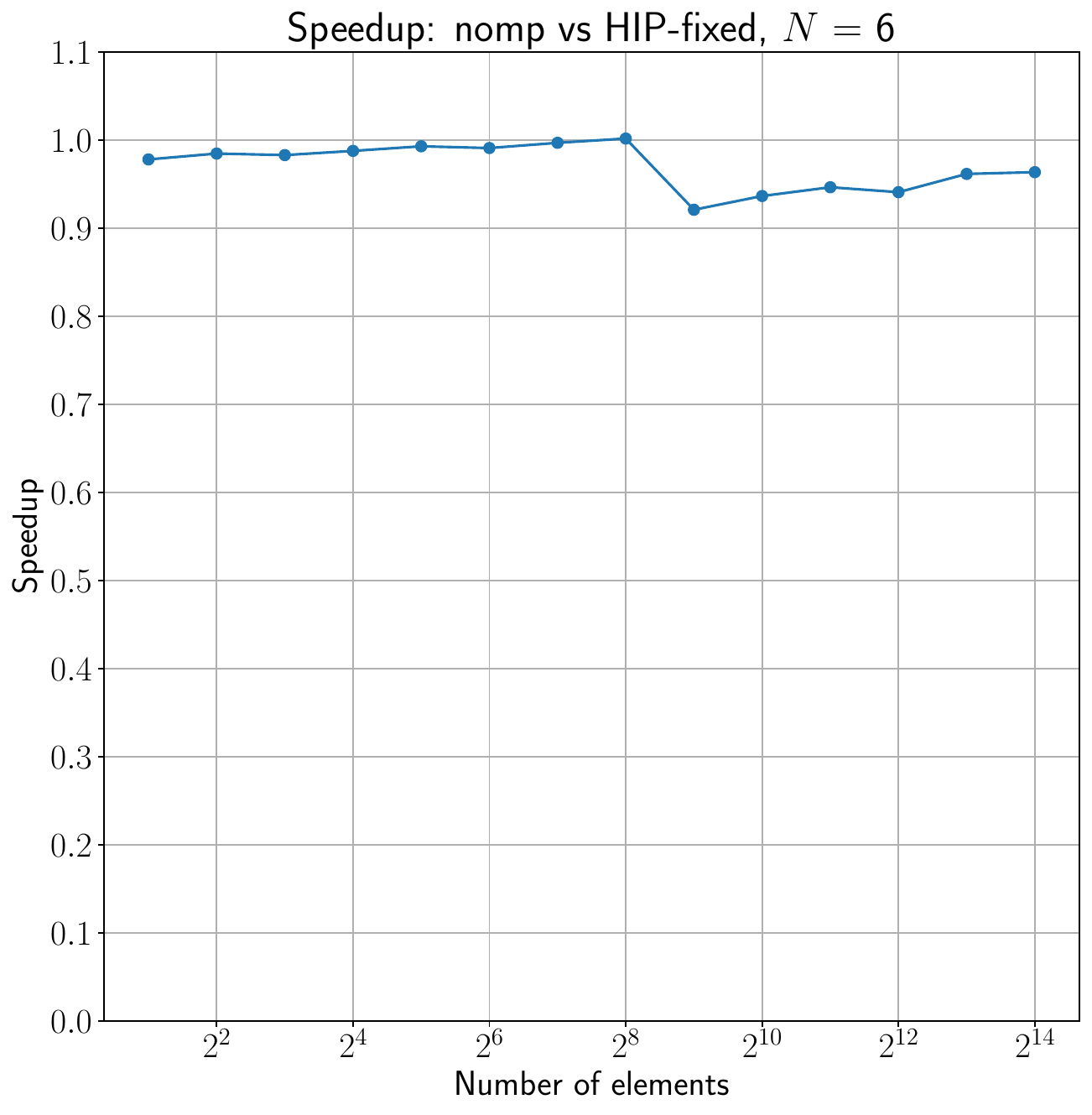}}}
\caption{\label{fig:nomp2_N_6} HIP-fixed vs. \nomp~Nekbone performance for $N=6$ on a
single GCD of AMD MI250X GPU on Frontier supercomputer at OLCF.}
\end{figure*}

\begin{figure*}
\centering
\subfloat[Final residual]{{\includegraphics[width=\figratio\textwidth]{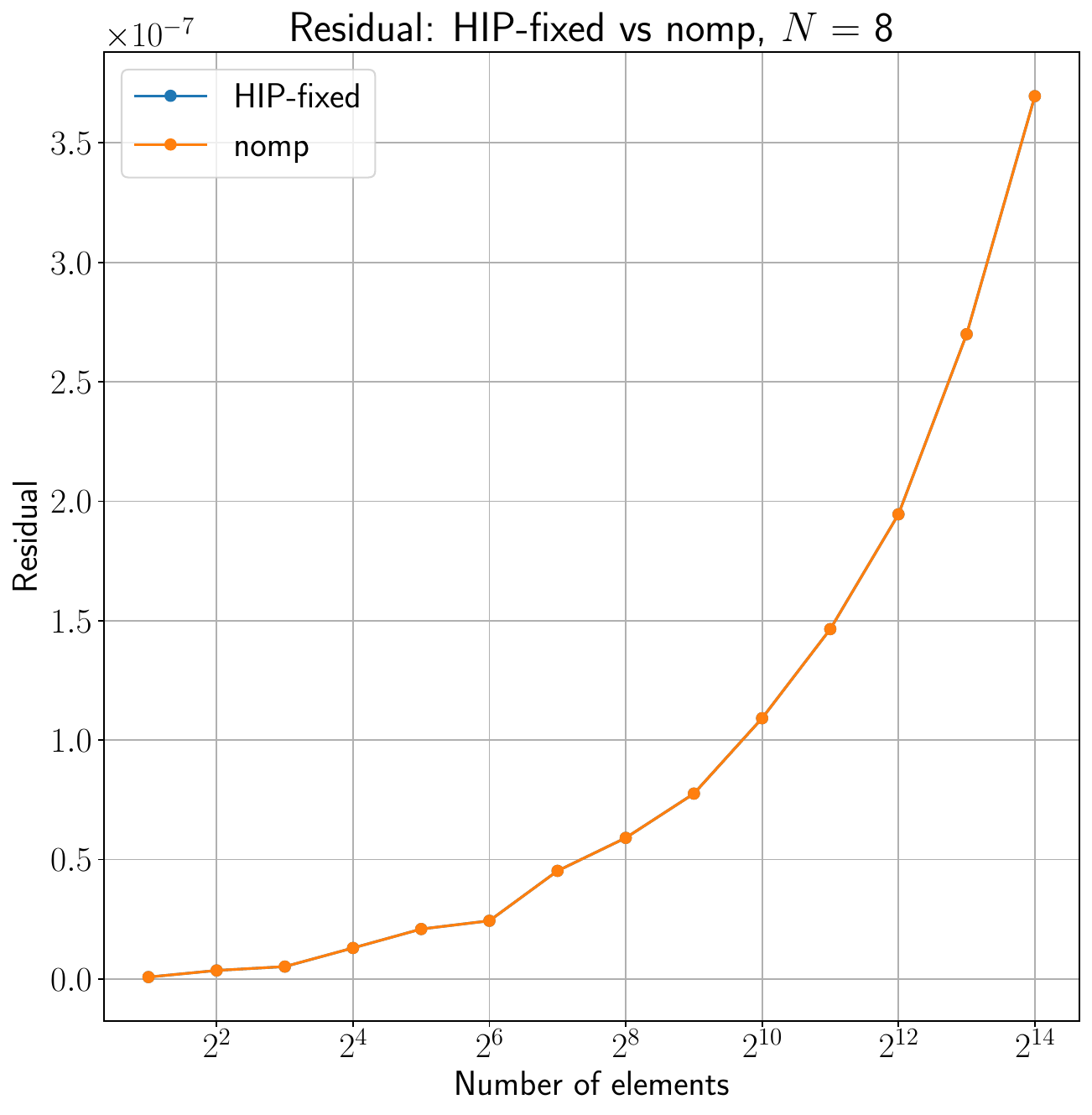}}}
\subfloat[Performance in GDOFS]{{\includegraphics[width=\figratio\textwidth]{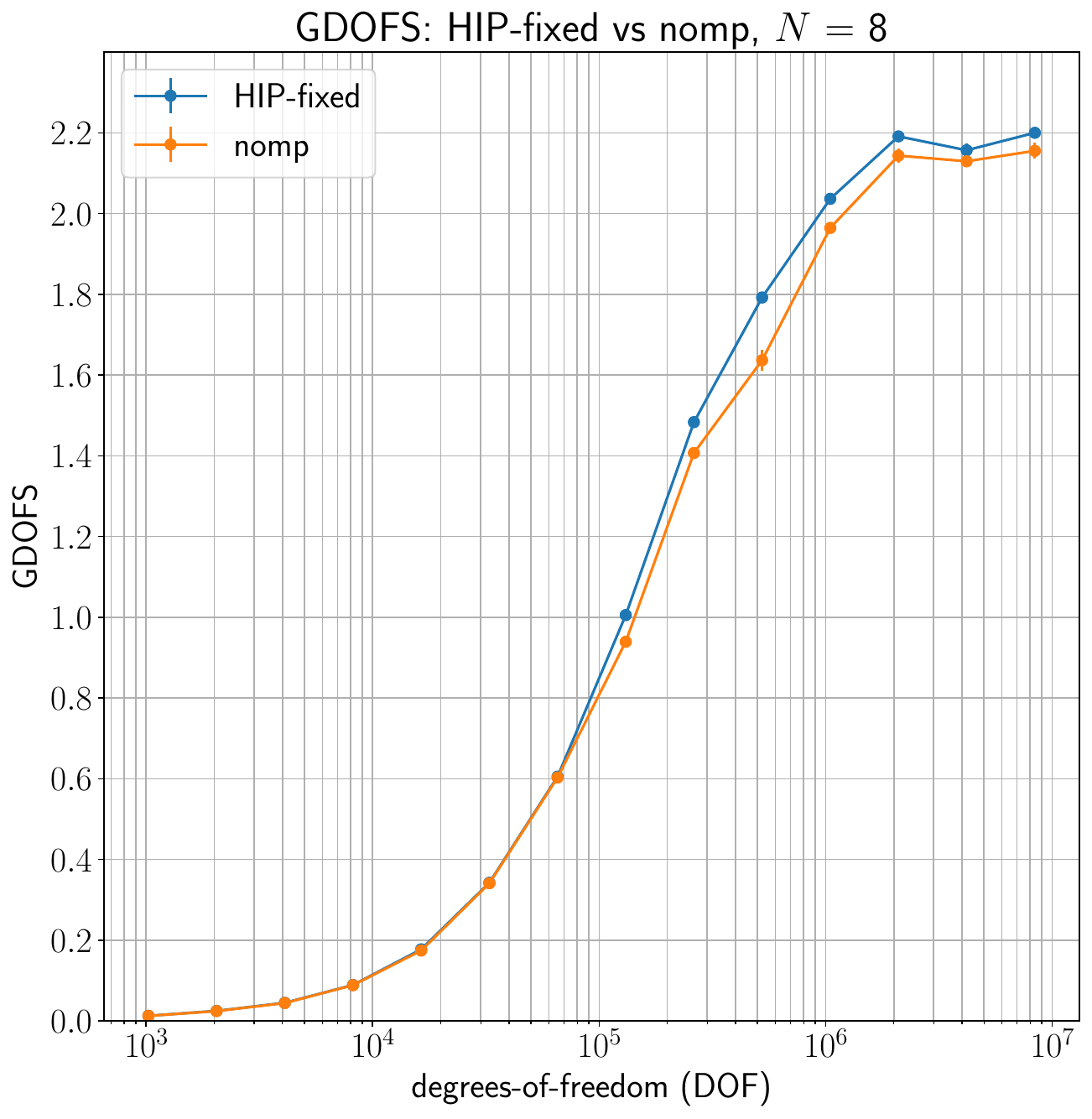}}}
\subfloat[Speedup]{{\includegraphics[width=\figratio\textwidth]{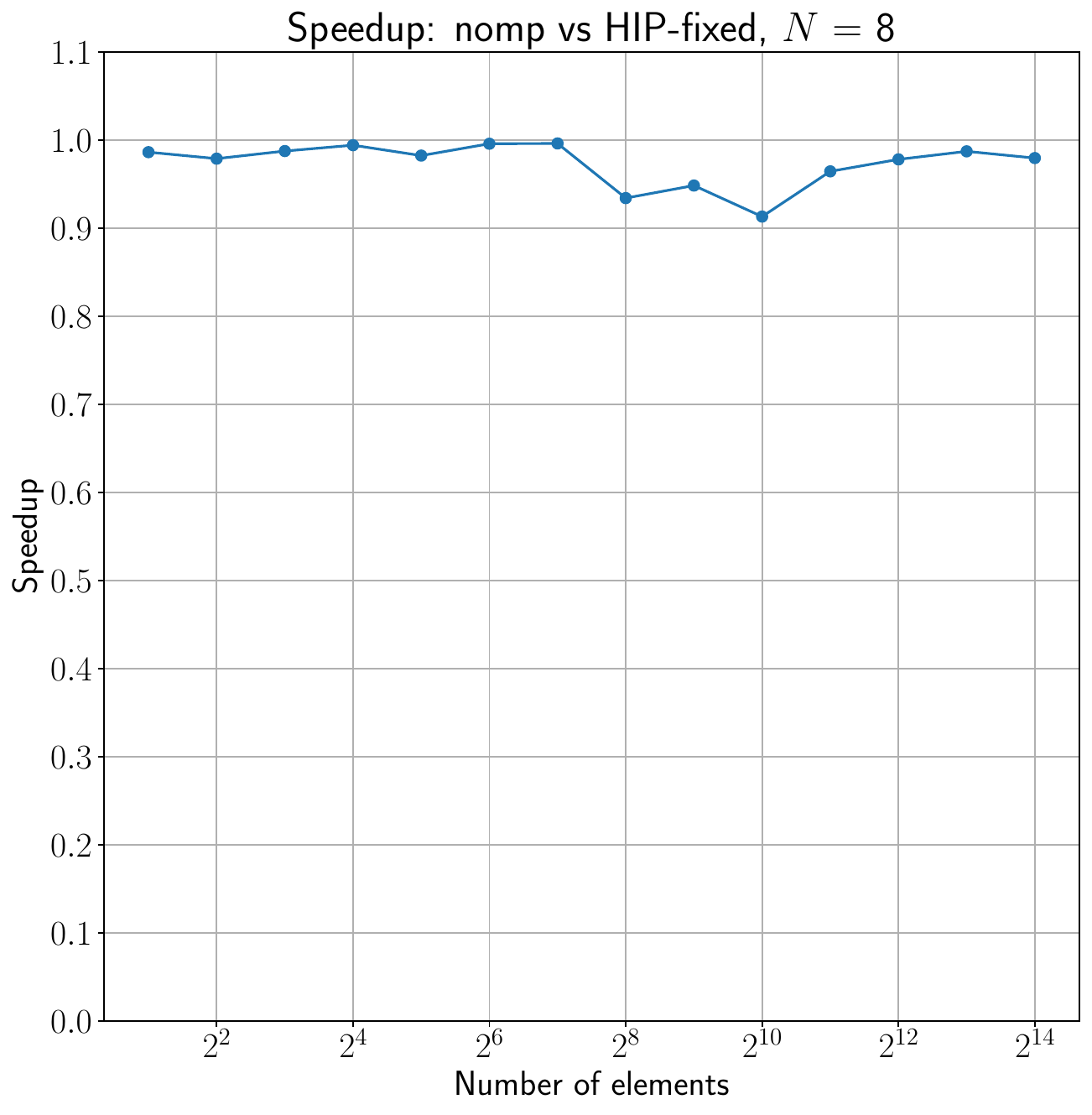}}}
\caption{\label{fig:nomp2_N_8} HIP-fixed vs. \nomp~Nekbone performance for $N=8$ on a
single GCD of AMD MI250X GPU on Frontier supercomputer at OLCF.}
\end{figure*}

\Cref{fig:nomp2_N_4} -- \Cref{fig:nomp2_N_8} compare \nomp~with the HIP-fixed
implementation (which includes he optimizations listed in
\Cref{tab:hip_vs_hip}).
HIP-fixed outperforms the \nomp~implementation marginally.
Since all the kernels are now identical between \nomp~and HIP-fixed
implementations, we expect HIP-fixed to outperform \nomp~as the former
does not have the function call overheads associated with the \libnomp~runtime
library. 
However, \nomp~still reach more than $90\%$ of the HIP-fixed implementation
performance across all the cases.

\subsection{Lines of code: \nomp~vs HIP}
\begin{table}\centering
\begin{tabular}{lc}
\toprule
\textbf{Implementation} & \textbf{Lines-of-code}\\
\midrule
\nomp & 345 (C: 266, Python: 79)\\
HIP-variable & 374\\
HIP-fixed & 526\\
\bottomrule
\end{tabular}
\caption{Lines-of-code comparison between \nomp, HIP-variable and HIP-fixed.}
\label{tab:loc}
\end{table}
We conclude this section with a lines-of-code (LOC) comparison between \nomp,
HIP-variable and HIP-fixed implementations.
While this metric can be subjective, it provides a crude estimate for the
programmer effort required for developing and maintaining the source code.
The LOC values for each of the backends we considered for performance
evaluation is shown in \Cref{tab:loc}.
We include both C Nekbone implementation and Python scripts containing
loopy transformation code in the case of \nomp.
\nomp~is able to achieve comparable performance to HIP-fixed while using
considerably lesser LOC.

\section{Conclusion}\label{sec:conclusion}
We presented \nomp: a framework for building domain specific compilers
using the C programming language.
We demonstrated that \nomp~improves programmer productivity by its virtue
of reusing optimizations while achieving comparable performance to an
optimized low-level programming model using the SEM/FEM miniapp Nekbone
as an example.
All our experiments were done on a single GCD and multiple GPU support
(through MPI~\cite{gls99}) is necessary for scalable high-performance
scientific computing applications.
No changes are necessary in \nompcc~in order to support MPI other than
pointing it to MPI header files and libraries.
However, several improvements are necessary in \libnomp~to make it MPI-aware.
We plan to do those improvements in \libnomp~and scale \nomp~based real
world scientific applications on state-of-the-art supercomputers in future.

\bibliographystyle{IEEEtran}
\bibliography{./bibs/emmd}

\end{document}